\documentclass[12pt]{article}
\usepackage[utf8]{inputenc}
\usepackage{ucs}
\usepackage{amscd,amsmath,amssymb,amsfonts,latexsym,mathrsfs,amsthm}
\usepackage{graphicx} 
\usepackage{setspace}
\usepackage{underscore}
\usepackage{graphics,epsfig}
\usepackage{sectsty}
\usepackage{graphics,epsfig}
\usepackage[english]{babel}
\usepackage{amsmath}
\usepackage{amsfonts}
\usepackage{amssymb,amsthm}
\usepackage{bm}
\usepackage{mathrsfs}
\usepackage{subfigure}
\usepackage[usenames]{color}
\usepackage{rotating}
\usepackage{multicol}
\usepackage{multirow}
\usepackage{mathtools}
\usepackage[table]{xcolor}

\usepackage{url}

\usepackage{footmisc}

\usepackage{url}
\usepackage{graphicx}

\usepackage{flushend}
\usepackage{verbatim}
\usepackage{tabularx}
\usepackage{multirow} 
\usepackage{arydshln}
\usepackage{bm}
\usepackage{subfigure}

\DeclareUnicodeCharacter{2212}{-}

\UseRawInputEncoding

\title{ 
Universal and Automatic Elbow Detection for Learning the Effective Number of Components in Model Selection Problems
}
\author{Eduardo Morgado$^*$, Luca Martino$^*$, Roberto San Mill{\'a}n-Castillo$^*$,  \\
 {\small$^*$ 
Universidad Rey Juan Carlos (URJC), Madrid, Spain.}}
\date{2022}

\begin{document}

\maketitle
\begin{abstract}

We design a Universal Automatic Elbow Detector (UAED) for deciding the effective number of components in model selection problems. The relationship with the information criteria widely employed in the literature is also discussed. The proposed UAED does not require the knowledge of a likelihood function and can be easily applied in diverse applications, such as regression and classification, feature and/or order selection, clustering, and dimension reduction. Several experiments involving synthetic and real data show the advantages of the proposed scheme with benchmark techniques in the literature.  
\end{abstract}

\section{Introduction}\label{sec:introduction}

Model selection is vast and one of the most relevant tasks in signal processing, statistics
and machine learning \cite{bishop2006pattern,llorente2020review,ding2018model,stoica2022monte}. It is the process of selecting a statistical model from a set of candidates. Model selection includes as special cases the following well-known sub-tasks: order selection (e.g., in polynomial functions or auto-regressive models \cite{stoica2004model}), variable selection \cite{bolon2013review}, dimension reduction \cite{ma2013review}, and clustering \cite{saxena2017review}, to name a few.
\newline
More specifically, in a large amount of research works from the most diverse fields, researchers and practitioners face a trade-off between the number of components/variables to consider in their analyses and the performance of the obtained results. Note that we use the term ``variables" as a general concept that can equivalently represent variables, features, or the number of clusters, depending on the nature of the considered problem. This trade-off occurs because increasing the number of variables taken into account in the analysis allows for better results, at the expense of obtaining a more complex model. In other words, the model performance and  the model complexity generate the so-called bias-variance trade-off. Therefore, in many applications, researchers must obtain the optimal number of components/variables addressing the aforementioned trade-off \cite{bishop2006pattern}.
\newline
{\bf Related works.} The solutions in the literature belong to different families and approaches. 
A first class of methods is formed by the {\it resampling techniques}, such as cross-validation or bootstrap, where the dataset is 
split into training and test sets \cite{Stoica04digital,fong2020marginal,vehtari2017practical}. However, the proportion of data to include in the training and test sets is a crucial parameter that affects critically the results.
Another important family is the class of the 
{\it information criteria} \cite{konishi2008information}, such as the  Bayesian Information Criterion (BIC) \cite{schwarz1978estimating}, the Akaike Information Criterion (AIC) \cite{Spiegelhalter02}, or the Hannan-Quinn Information Criterion (HQIC) \cite{Hannan79}, to name a few
\cite{llorente2020review,SafePriorsLlorente}. The information criteria consider a linear penalization of the model complexity, and they differ for the choice of the slope of this penalization. These choices are motivated by theoretical probabilistic derivations which involve several assumptions and approximations. Hence, the good performance of an information criterion is often restricted to very specific scenarios.  
Moreover, the computation of the information criteria often requires the knowledge of the maximum of a likelihood function. More recent and advanced works related to information criteria can be found in \cite{Drton17,Mariani15,SIC}.
Other probabilistic strategies related to the information criteria are the so-called  minimum description length principle,  Mallows's Cp coefficient and the structural risk minimization \cite{kobayashi1990mallows,shawe1998structural}. In the Bayesian framework, the use of marginal likelihood and posterior predictive approaches are usually employed 
\cite{llorente2020review,BMA99,Pooley18}. The connection between the marginal likelihood and information criteria is discussed in the appendices of \cite{SafePriorsLlorente}. The posterior predictive approach is related to the cross-validation idea.
Furthermore, standard frequentist approaches based on $p$-values have a vast use in some specific applications and deserve to be cited \cite{efroymson1960multiple,hocking1976biometrics}. Finally, some authors apply a visual inspection
of an error curve looking for an ``elbow", specially in the clustering literature.
\newline
In this work, we design a Universal Automatic Elbow Detector (UAED) based on a geometric approach. The proposed scheme is inspired by the concept of the maximum ``area under the curve'' in receiver operator characteristic curves \cite{bishop2006pattern,HanleyAUC82}, which is well-known and vastly employed in signal processing and machine learning. The resulting UAED technique also induces a linear penalization of the model complexity.  We discuss the connections, differences and the advantages of UAED with respect to the information criteria already presented in the literature. It is important to remark that the range of applicability of UAED is much wider than other techniques in the literature, since no likelihood function is needed. The application of UAED only requires the knowledge of an error curve, that can be defined in different ways according to the user's desire. 
Moreover, we describe several appealing behaviors of UAED and test it in different numerical examples, three of them involving real datasets. The results show the benefits of UAED with respect to other benchmark techniques in the literature. Therefore, the main contributions of the work are the following:
\begin{itemize}
\item We introduce a Universal Automatic Elbow Detector (UAED) based on a geometric approach.
\item Four equivalent derivations of UAED are presented, two of them in Section \ref{sub:Calculo} and other two derivations in Appendices \ref{GrandeLucaEdu}-\ref{otherder}. 
\item The behavior, invariance and other properties of UAED are discussed. See Section \ref{SuperSectBeh} and Appendix \ref{AppSuperLuca}.
\item UAED is tested and compared with other benchmark methods in six numerical experiments, regarding different applications (clustering, order selection and variable selection).  Three of them involve applications with real datasets. See Section \ref{sec:Casos_Prac}. 
\item We provide a related Matlab code of UAED.\footnote{A related Matlab code is given at \url{http://www.lucamartino.altervista.org/PUBLIC_UAED_CODE.zip}.}
\end{itemize}
\noindent
The rest of the article is organized as follows. Section \ref{sub:Ini_Curve} describes the framework and the notation employed in the derivation of UAED. Section \ref{sub:Calculo} presents and discusses UAED in detail, whereas Section \ref{sec:Casos_Prac} shows six numerical experiments and practical applications (three experiments involve real datasets). Finally, in Section \ref{sec:conclusions}, some conclusions are given. In the appendices, we present (a) two alternative derivations, (b) an additional property of UAED, and (c) a possible extension of UAED (which provides more flexibility).

\section{Framework and main notation}\label{sub:Ini_Curve}

\begin{figure}[!ht]
   \centerline{
    \subfigure[\label{Fig1Luca}]{\includegraphics[width=0.5\columnwidth]{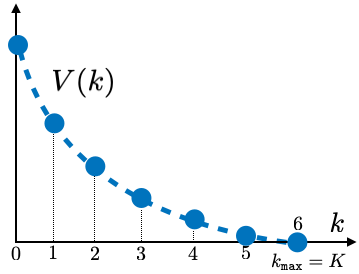}}
        \subfigure[\label{Fig0Luca}]{\includegraphics[width=0.5\columnwidth]{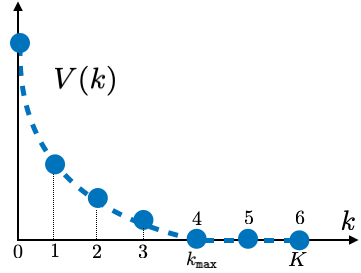}}
       }
      \centerline{
          \subfigure[\label{Fig2Luca}]{\includegraphics[width=0.5\columnwidth]{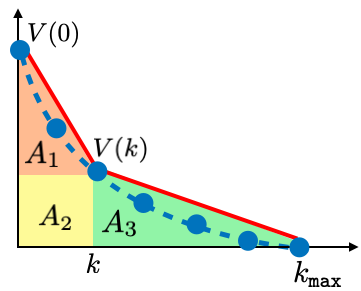}}
  }
   \caption{{\bf (a)}-{\bf (b)} Example of error curve $V(k)$ where {\bf (a)} $k_{\texttt{max}}=K=6$, {\bf (b)} $k_{\texttt{max}}=4$ and $K=6$.  {\bf (c)} Construction with two straight lines and the areas $A_1$, $A_2$ and $A_3$.}
   \label{fig:Curva_Inicial}
\end{figure}

In many applications, we desire to infer a vector of parameters ${\bm \theta}_k=[\theta_1,...,\theta_k]^{\top}$ of dimension $k$ given a data vector ${\bf y}=[y_1,...,y_N]^{\top}$. A likelihood function $p({\bf y}| {\bm \theta}_k)$ is usually available, often induced by a related physical model. Furthermore, in different types of real-world application problems (clustering, variable selection, or dimension reduction) and specially in model selection problems,  an {\it error function} (i.e., a fitting measure) is obtained. Here, we denote it as
$$
V(k): \mathbb{N} \rightarrow \mathbb{R}, \qquad k=0,1,2,...,K,
$$
where $k$ denotes the number of components (e.g., variables, clusters, or order of the polynomial function), i.e., $k$ defines the complexity of the model. In the literature, we often have
$$
V(k)=-2\log(\ell_{\texttt{max}}), \quad \mbox{ where } \quad \ell_{\texttt{max}}=\max_{\bm{\theta}} p({\bf y}| \bm{\theta}_k),
$$ 
as in \cite{konishi2008information}. However, in this work, $V(k)$ could be directly the mean square error (MSE), or the mean absolute error (MAE).  For instance, $V(k)$ can represent the prediction error in regression problems with a polynomial function, where $k$ is the order of the polynomial, or the sum of the inner variances within clusters where $k$ is the number of clusters. We assume that $k$ starts in $0$ and grows with step 1 for simplicity, but more general cases can easily be addressed. Namely, a different incremental step could be also considered. 
\newline
Generally, $V(k)$ is a {\it non-increasing} error curve, i.e., for any pair of non-negative integers $n_1,n_2$ such that $n_2 > n_1$, then we have $V(n_2) \leq V(n_1)$.\footnote{This condition could be also relaxed. We keep it, for the sake of simplicity.} Indeed, $V(k)$ is a fitting term that decreases as the complexity of the model (given by the number $k$ of parameters) grows. Therefore, we have $V(0)\geq V(k), \forall k$. Hence, in this work, we assume $V(k)$ to be a non-increasing function. See Figure \ref{fig:Curva_Inicial} for some graphical examples.
\newline
Observe that $V(0)$ represents the value of the error function corresponding, for instance, to a constant model in a regression problem, or a single cluster (for all the data) in a clustering problem.
 In some applications, the score function $V(k)$ should be also convex, i.e., the differences $V(k+1)-V(k)$ will decrease as $k$ increases. This is the case of a variable selection problem, if the variables have been ranked correctly.  However, this work does not require conditions regarding the concavity of $V(k)$. 
\newline
\newline
{\bf Additional assumptions.} Just for the sake of simplicity and without loss of generality, we assume that $\min V(k)=V(K)=0$. Note that this condition can be always obtained with a simple subtraction, defining a new curve 
\begin{equation}\label{ImpEqquitaMin}
V'(k)=V(k)-\min V(k)=V(k)-V(K).
\end{equation} 
Figure \ref{Fig1Luca} depicts a graphical example of the curve $V(k)$.
Moreover, above we have assumed $k=0,1,...,K$ but, if there exists a value $k_{\texttt{max}} \leq K$ such that $V(k)$ has not an additional drop for $k\geq k_{\texttt{max}}$, i.e., $k_{\texttt{max}}=\min\left[\arg\min_k V(k)\right]$, so that
\begin{align}
V(k_{\texttt{max}})=V(k_{\texttt{max}}+1)=V(k_{\texttt{max}}+2)=...=V(K)=0.
\end{align}
In this scenario, we can consider $k=0,1,..., k_{\texttt{max}}$, since the rest of the components, from $k_{\texttt{max}}+1$ to $K$, must be discarded because they do not cause a drop in the error function. See Figures  \ref{Fig1Luca}-\ref{Fig0Luca} for two graphical examples. Clearly,
if the minimum value of $k$ is different from $0$, let us say $k_{\texttt{min}}$,  we can always set $k'=k-k_{\texttt{min}}$.   

\section{The Universal Automatic Elbow Detector (UAED)}\label{sub:Calculo}

In this section, we provide two equivalent geometric derivations of the proposed method, and discuss the similarities, differences, and  connections with other methods in the literature. 
The behavior of the proposed technique is  described and some interesting considerations are also highlighted.

\subsection{First derivation}\label{FirstDerSect}
Considering the decay $V(k)$ described in the previous section, the underlying idea is ``inspired" by the concept of the maximum AUC in ROC curves \cite{bishop2006pattern,HanleyAUC82}. Namely, we desire to extract geometric information from the curve $V(k)$ looking for an ``elbow'' in order to determine the optimal number of components, denoted $k^*\in \{0,1...,k_{\texttt{max}}\}$, to include in our model (i.e., in the vector $\bm{\theta}_{k^*}$). 
\newline
We consider the construction of two straight lines passing through the points    
$(0,V(0))$, $(k, V(k))$ and $(k, V(k))$, $(k_{\texttt{max}},0)$ as shown in Figure \ref{Fig2Luca} (where $k\in\{0,1,...,k_{\texttt{max}}\}$). These two  straight lines form a piece-wise linear approximation of the curve $V(k)$. The goal is to minimize  
 the area under this approximation. 
More specifically, as we can see in Figure~\ref{Fig2Luca}, the area to minimize consist of three sub-areas: the two areas of two triangles ($A_1$ and $A_3$) and the area of a rectangle in the middle ($A_2$). Namely, we have 
\begin{align}
A_1 &= \frac{k \cdot (V(0) - V(k))}{2}, \\
A_2 &= k \cdot V(k),\\ 
A_3 &=\frac{(k_{\texttt{max}} - k) \cdot V(k)}{2}, 
\end{align}
hence the definition of $k^{*}$ is
\begin{align}
k^{*} = \arg\min_k\{A_1 + A_2 + A_3\}&= \arg\min_k\left\{ \frac{k(V(0) - V(k))}{2} + k  V(k) + \frac{(k_{\texttt{max}} - k) V(k)}{2}\right\}, \nonumber \\
&= \arg\min_k\left\{ \frac{kV(0)}{2}  + \frac{k_{\texttt{max}} V(k)}{2}\right\}, \nonumber \\
&= \arg\min_k\left\{ \frac{V(k)}{V(0)}+\frac{k}{k_{\texttt{max}}}  \right\}, \quad \mbox{for $k=1,...,k_{\texttt{max}}$}, \label{exp:N_opt}
\end{align}
where clearly we are assuming $V(0)\neq 0$ and $k_{\texttt{max}}\neq 0$. Note that in the last equation we have multiplied by the constant factor $\frac{2}{V(0)k_{\texttt{max}}}$. Now, multiplying the last expression by the constant value $V(0)$, we can equivalently write 
\begin{equation}\label{exp:N_opt_2}
k^{*} = \arg\min_k\left\{ V(k)+\frac{V(0)}{k_{\texttt{max}}} k \right\},  \quad \mbox{for $k=1,...,k_{\texttt{max}}$}. 
\end{equation}
It is important to remark that, since $k$ belongs to a discrete and finite set, solving the optimization above is straightforward. In the case of multiple minima, e.g., having $M$ different minima,
$ k^{*}_1, k^{*}_2,..., k^{*}_M$, the user can choose the best solution (within the $M$ possible one) according to some specific requirement depending on the specific application. Here, we suggest the most conservative choice, i.e., 
\begin{equation}\label{EqMaxMin}
    k^*=\max k^{*}_j,  \quad \mbox{for $j=1,...,M$}.
\end{equation}
\subsection{Second equivalent derivation}\label{GrandeEdu}

\begin{figure}[!ht]
   \centerline{
 \includegraphics[width=0.5\columnwidth]{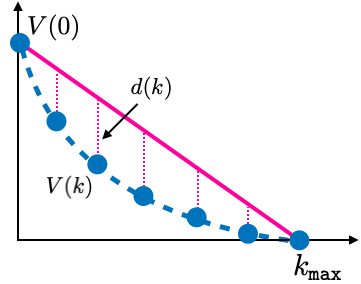}   
  }
   \caption{ Graphical representation of alternative derivation in Section \ref{GrandeEdu}.}
   \label{Fig3Luca}
\end{figure}

 The solution offered by the expressions \eqref{exp:N_opt}-\eqref{exp:N_opt_2} is equivalent to finding the $k^{*}$ such that
  the difference between $V(k^{*})$ and the value of the straight line (evaluated at $k^*$, as well) connecting the extreme points $(0,V(0))$ and $(k_{\texttt{max}},0)$ is maximized, as depicted in Figure \ref{Fig3Luca}. More specifically, this straight line has an equation
 \begin{align}
 v(k)=-\frac{V(0)}{k_{\texttt{max}}}\cdot k+V(0),
 \end{align}
 hence, the difference that we maximize is the following:
 \begin{align}
 d(k)&=v(k)-V(k), \nonumber \\  
 &=-\frac{V(0)}{k_{\texttt{max}}}\cdot k+V(0)-V(k), \nonumber \\
 &=V(0)-\left(\frac{V(0)}{k_{\texttt{max}}}\cdot k+V(k)\right).
 \end{align}
Since $V(0)$ does not depend on $k$ (i.e., it is a constant value), we can write
 \begin{align}
 k^*=\arg \max_k d(k)&=\arg \max_k\left[V(0)-\left(\frac{V(0)}{k_{\texttt{max}}}\cdot k+V(k)\right)\right], \nonumber \\
 &=\arg \max_k\left[-\left(\frac{V(0)}{k_{\texttt{max}}}\cdot k+V(k)\right)\right], \nonumber \\
 &=\arg \min_k\left[\frac{V(0)}{k_{\texttt{max}}}\cdot k+V(k)\right],
 \end{align}
which is exactly the expression in Eq. \eqref{exp:N_opt_2}.
Two additional and equivalent derivations are given in Appendix \ref{GrandeLucaEdu} and Appendix \ref{otherder}. They are also represented graphically in Figures \ref{Fig4Luca} and \ref{figLucaMuchas_2}, respectively.
 
\subsection{Relation with the information criteria}
Recalling the expression in \eqref{exp:N_opt_2}, i.e.,
\begin{equation*}
k^{*} = \arg\min_k\left\{ V(k)+\frac{V(0)}{k_{\texttt{max}}} k \right\},
\end{equation*}
here we show that this cost function can be interpreted in the same form of other information criteria, i.e.,  with a linear penalization of the model complexity,
\begin{align}\label{HereNewBIC}
C(k)&=V(k)+\frac{V(0)}{k_{\texttt{max}}} k,  \\
&=V(k)+\lambda k, \label{ICeq}
\end{align}
where we set $\lambda=\frac{V(0)}{k_{\texttt{max}}}$. Note that Eq. \eqref{ICeq} has exactly the same form of the cost function used in the information criteria like BIC and AIC, for instance, when $V(k)$ is defined as
$$
V(k)=-2\log \ell_{\texttt{max}}, \quad \mbox{ with } \quad \ell_{\texttt{max}}=\max_{\bm{\theta}} p({\bf y}| \bm{\theta}_k).
$$
 BIC corresponds to the choice $\lambda=\log(N)$ where $N$ is the number of data in ${\bf y}$, and AIC corresponds to the choice $\lambda=2$.   
Therefore, when $V(k)=-2\log \ell_{\texttt{max}}$, UAED can be interpreted as an information criterion with the particular choice of $\lambda=\frac{V(0)}{k_{\texttt{max}}}$. Table \ref{TablaIC} summarizes this information.

\begin{table}[!h]	
	\caption{Different information criteria  and the proposed UAED.}\label{TablaIC}
	\vspace{-0.3cm}
	\begin{center}
		\begin{tabular}{|c|c|} 
			\hline 
			{\bf Criterion} & {\bf Choice of } $\lambda$   \\ 
			\hline 
			\hline 
				&\\
			Bayesian-Schwarz Information Criterion (BIC) \cite{schwarz1978estimating} &  $\log N$ \\
			&\\
			Akaike Information Criterion  (AIC) \cite{Spiegelhalter02} &  $2$ \\
			&\\
			Hannan-Quinn Information Criterion (HQIC)  \cite{Hannan79}&  $\log(\log(N))$ \\
		&\\		Universal Automatic Elbow Detector (UAED)&  $\frac{V(0)}{k_{\texttt{max}}}$ \\
			& \\
			\hline
		\end{tabular}
	\end{center}
\end{table}
\subsection{Behavior of the proposed solution}\label{SuperSectBeh}

Analyzing the involved parameters in the expression \eqref{exp:N_opt_2} or \eqref{HereNewBIC}, we can highlight the following considerations about the behavior of the UAED method. We list some important points below:
\newline
\newline
$\bullet$ Observing Eq. \eqref{HereNewBIC}, the penalization of the complexity of the model depends on $V(0)$ and $k_{\texttt{max}}$: since $\lambda=\frac{V(0)}{k_{\texttt{max}}}$  increasing $V(0)$ or decreasing $k_{\texttt{max}}$, intensifies the penalty. This is a reasonable and desirable behavior. Indeed, increasing the value of $V(0)$ also increases the differences $V(0)-V(k)$, which means that the first components/variables have more impact in the fitting - the decay of $V(k)$ - so that fewer components/variables can form a reasonable model. Otherwise, decreasing the value of $V(0)$ means more variables have a similar impact in the decay of $V(k)$. Therefore, we should consider more components, in fact the slope of the penalization, $\lambda=\frac{V(0)}{k_{\texttt{max}}}$, decreases in this case.
\newline
$\bullet$ Regarding $k_{\texttt{max}}$, we can notice that a decrease of $k_{\texttt{max}}$ means that fewer components/variables  produces a drop in the curve $V(k)$.  On the other hand, an increase in $k_{\texttt{max}}$ means that the use of more variables causes a drop $V(k)$, so we should consider more components, indeed, the slope of the penalization, $\lambda=\frac{V(0)}{k_{\texttt{max}}}$, decreases.
\newline
$\bullet$ {\bf Scaling of the axes.} Looking the expression \eqref{exp:N_opt_2} or \eqref{HereNewBIC}, it is possible to show that the solution does not depend on different possible re-normalization of the axes, i.e., after a scaling of the axes (one of them, or both, even with different scales) the solution remains invariant, or is a scaled version of the previous one (with the same scaling factor). Indeed, considering a scaling on the vertical axis, i.e., assuming $V(k)'=a V(k)$ with $a>0$, we have
\begin{align}
k^{*} = \arg\min_k\left[ aV(k)+\frac{aV(0)}{k_{\texttt{max}}} k \right]&= \arg\min_k \left[a \left( V(k)+\frac{V(0)}{k_{\texttt{max}}} k \right)\right], \nonumber \\
&=\arg\min_k\left[ V(k)+\frac{V(0)}{k_{\texttt{max}}} k \right].
\end{align}
Let now consider the case of scaling the horizontal axis, for instance, instead of having $k=0,1,2...,k_{\texttt{max}}$, we have $k'=0,b,2b,...,bk_{\texttt{max}}$ (i.e., $k'=b k$), and another error curve defined as $\widetilde{V}(k')=V(k'/b)$, where  $b$ is a positive integer. Hence we can write
 \begin{align}
 (k')^{*} = \arg\min_{k'}\left[ \widetilde{V}(k')+\frac{V(0)}{k_{\texttt{max}}'} \cdot k' \right]&= \arg\min_{k'}\left[ V(k'/b)+\frac{V(0)}{k_{\texttt{max}}'} \cdot k' \right], \nonumber \\
 &=b\arg\min_{k}\left[ V(bk/b)+\frac{V(0)}{bk_{\texttt{max}}} \cdot bk \right], \nonumber \\
 &=b\arg\min_{k}\left[ V(k)+\frac{V(0)}{k_{\texttt{max}}} \cdot k \right]=bk^*.
 \end{align}
 Namely, the new solution $(k')^{*}=bk^*$ is just a scaled version of the previous one, taking into account the same scaling factor $b$.
 \newline
 $\bullet$ {\bf Shift of the axes.} Let us consider a $V(k)$ which fulfills that assumptions provided above (to be non-increasing and $\min V(k)=V(K)=0$). A shift of the vertical axis, i.e., $\widetilde{V}(k)=V(k)+c$ where $c\in \mathbb{R}$ does not affect the results since, by assumption, we have always to consider an error curve such that
$$
V'(k)=\widetilde{V}(k)-\min \widetilde{V}(k)=V(k)+c-c=V(k).
$$
 then $V'(k)$ satisfies again $\min V'(k)=0$. A shift on the horizontal axis only produces the same shift in the solution $k^*$.
Furthermore, given the considerations in Appendices \ref{GrandeLucaEdu} and \ref{otherder}, we can see that the solution $k^*$ is invariant even if the axes are exchanged.  
\newline 
$\bullet$ Here, we describe two ideal scenarios and discuss the behavior of UAED. For clarity in the exposition, let us consider as an example a variable selection problem. First of all, we consider the case that all the input variables are equally important for predicting the output variable. Then, we have $k_{\texttt{max}}=K$, and 
the error curve $V(k)$ is a straight line connecting the points $(0,V(0))$ and $(k_{\texttt{max}},V(k_{\texttt{max}}))$ (i.e., each variable has the same impact to the error decay). In this scenario, we have $k_{\texttt{max}}=K$, and UAED provides $M=k_{\texttt{max}}+1$ different minima
$ k^{*}_1=0, k^{*}_2=1, k^{*}_3=2,..., k^{*}_M=k_{\texttt{max}}$ (i.e., $M$ possible candidates to be an elbow). Thus, the UAED solution is given by Eq. \eqref{EqMaxMin}, i.e.,  $k^*=\max k^{*}_j=k_{\texttt{max}}$. Namely, UAED suggests to select all the variables, that is the correct solution.
\newline
On the other hand, let us consider now a scenario where all the input variables are independent from the output variable. In this case, $V(k)$ is a constant function, i.e., $V(k)=V(0)$ for all $k$ and, as a consequence, $k_{\texttt{max}}=0$. Hence, since $k_{\texttt{max}}=0$, UAED gives $k^*=0$, which is the correct solution (i.e., no variables should be selected). Thus, in both scenarios, UAED provides the correct results.

\section{Experiments with synthetic and real data}\label{sec:Casos_Prac}

In this section, we test the UAED in six real-world applications.
In each experiment, we consider a different function $V(k)$, in order to show the vast range of applicability of UAED. Sections \ref{CluSect}, \ref{PolOrder}, \ref{ARexample} consider synthetic data in a clustering example and two order selection problems. In Sections \ref{VSwithRealData}, \ref{VSwithOscar} and \ref{VS_BIO_2}, the experiments involve the analysis of real datasets: the first one is a variable selection in a regression problem with soundscape emotion data,  whereas the second and third ones involve classification problems with real biomedical datasets. We compare the performance of UAED with BIC, AIC, and other information criteria described in the literature, in those examples where these schemes can be also applied.




\subsection{Clustering}\label{CluSect}
\begin{figure}[h!]
\centerline{
\subfigure[\label{Fig4Clu_a}]{\includegraphics[width=9cm]{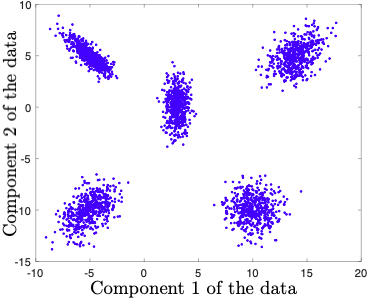}}
\subfigure[\label{Fig4Clu_b}]{\includegraphics[width=9cm]{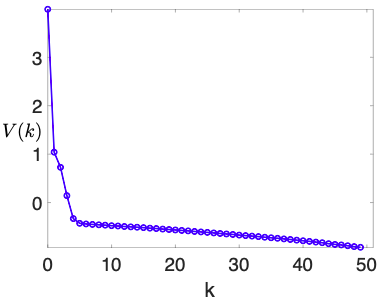}}
}
\caption{\footnotesize {\bf (a)} Artificial Data  of the clustering experiment. {\bf (b)} The function $V(k)=\log\left[\sum_{j=1}^{k+1} \mbox{var}(j)\right]$ where $\mbox{var}(j)$ represents the inner variance in the $j$-th cluster. Note that $k=0$ corresponds to a unique, single cluster. 
}
\label{Fig4Clu}
\end{figure}

We consider $2500$ simulated data from a mixture of $5$  bidimensional Gaussian distributions, $\mathcal{N}({\bm \mu}_i, {\bf \Sigma}_i)$, where ${\bm \mu}_1=[3,0]$, ${\bf \Sigma}_1=[0.3,0;  0, 2]$,  ${\bm \mu}_2=[14,5]$, ${\bf \Sigma}_2=[1.5,0.7;0.7,1.5];$, ${\bm \mu}_3=[-5,-10]$, ${\bf \Sigma}_3=[1.5,0.7;0.7, 1.5]$, ${\bm \mu}_4=[10,-10]$, ${\bf \Sigma}_4=[1.5,0;0,1.5];$, and ${\bm \mu}_5=[-5,5]$, ${\bf \Sigma}_5=[1,-0.8;-0.8,1]$. Figure \ref{Fig4Clu_a} shows these data points.
\newline
We assume $V(k)=\log\left[\sum_{j=1}^{k+1} \mbox{var}(j)\right]$, where $\mbox{var}(j)$ represents the inner variance of the $j$-th cluster, as shown in Figure \ref{Fig4Clu_b}. Each value of $\mbox{var}(j)$ has been computed and averaged over $200$ runs, applying a k-means algorithm. In this setting, the total number of clusters is given by $k+1$ (i.e., $k=0$ corresponds to a unique, single cluster). We assume $K=50$ as the maximum number of possible clusters.
\newline
It is important to remark that, with this choice of $V(k)$, the other information criteria cannot be directly applied\footnote{\label{footnoteHere}The information criteria require the choice of the error curve of type $V(k)=-2\log \ell_{\texttt{max}}$ where $\ell_{\texttt{max}}=\max_{\bm{\theta}} p({\bf y}| \bm{\theta}_k)$ and, as a consequence, a definition of a likelihood function $p({\bf y}| \bm{\theta})$.}. We apply UAED and obtain  $k^*+1=5$ as the chosen number of clusters, which is the correct solution. 

\subsection{Order selection of a polynomial function in a regression problem}\label{PolOrder}
 We generate a dataset of $N=100$ pairs  $\{x_n,y_n\}_{n=1}^N$,  where  both inputs $x_n$'s and outputs  $y_n$'s are scalar values, considering the following observation model,
 \begin{align}\label{aquiPol}
    y_n&=\theta_0+\theta_1 x_{n}+\theta_2 x_{n}^2+...\theta_k x_{n}^k+ \epsilon_n,  
\end{align}
 where ${\bm \theta}_k=[\theta_0, \theta_1,...,\theta_k]^{\top}$,  $ \epsilon_n$ is Gaussian noise with zero mean and variance $\sigma_\epsilon^2=1$. The dataset has been generated with a polynomial function of order $k^*=4$, and with the coefficients
 $$
 \theta_0=4.05,\mbox{ }\theta_1=-2.025,\mbox{ }\theta_2=-2.225,\mbox{ }\theta_3=0.1,\mbox{ }\theta_4=0.1.
 $$
  In this experiment, we consider  $V(k)=-2\log(\ell_{\texttt{max}})$ with  $\ell_{\texttt{max}}=\max_{\bm \theta} p({\bf y}| {\bm \theta}_k)$ with $k\leq K$, where $p({\bf y}| {\bm \theta}_k)$ is induced by Eq. \eqref{aquiPol}, in order to allow the comparison with other schemes in the literature, as shown in Table \ref{TablaIC}. The corresponding function $V(k)$ is shown in Figure \ref{orderPolLabel}.  
\newline
\newline
Applying BIC, AIC and HQIC 
we obtain the suggested order of polynomial is 4, 6, and 10, respectively. With the proposed UAED method, we obtain the suggested  order is 4, which is the correct order of the underlying polynomial function.  Therefore, in this experiment,  BIC and UAED provide the correct answer.

\subsection{Order selection in an auto-regressive model}\label{ARexample}
We generate a dataset of $T$ pairs $\{t,y_t\}_{t=1}^T$, where $t$ is an integer temporal index and the signal  $y_t$  is a scalar value for each $t$. We consider the following auto-regressive model,
 \begin{align}\label{aquiAR}
    y_t&=\theta_1y_{t-1}+\theta_2 y_{t-2}+...\theta_{k} y_{t-k}+ \epsilon_t,   \quad \mbox{for $t=1,...,T,$}
\end{align}
 where ${\bm \theta}_k=[\theta_1, \theta_2,...,\theta_{k}]^{\top}$,  $ \epsilon_t$ is Gaussian noise with zero mean and variance $\sigma_\epsilon^2$. The order of the model is $k$. 
\newline
In this example, we consider two possible values of the order of the model, $k^*\in\{3,5\}$ and we have set $k_{\max}=K=100$. We generate the data ${\bf y}=[y_1,...,y_T]$ according to the model \eqref{aquiAR} considering the following coefficients
 \begin{align*}
& \theta_1=1,\mbox{ }\theta_2=-0.7408,\mbox{ }\theta_2=0.5488,\mbox{ }\theta_3=0.1, \qquad \qquad \qquad \qquad \qquad \qquad \qquad \mbox{ }\mbox{for $k=3$,} \\
 & \theta_1=1,\mbox{ }\theta_2=-0.7408,\mbox{ }\theta_2=0.5488,\mbox{ }\theta_3=0.1,\mbox{ }\theta_4=-0.4066, \mbox{ }\theta_5= 0.3012, \qquad \mbox{for $k=5$,}    
  \end{align*}
where we have used the formula $\theta_i=(-1)^{i-1}\exp\{-0.3(i-1)\}$, which ensures that the system in Eq. \eqref{aquiAR} is stable. 
We test different number of data, $T\in\{200, 2000\}$, and different levels of noise, with standard deviation $\sigma_\epsilon \in\{0.5,1,2\}$.
In all scenarios, we average the results with $10^3$ independent runs where, for each simulation, we generate a new time series of $T$ data according to Eq. \eqref{aquiAR}.  
\newline
Moreover, in all the simulations, we consider  $V(k)=-2\log(\ell_{\texttt{max}})$ with  $\ell_{\texttt{max}}=\max_{\bm \theta} p({\bf y}| {\bm \theta}_k)$ with $k\leq K$, where $p({\bf y}| {\bm \theta}_k)$ is induced by Eq. \eqref{aquiAR}, in order to allow the comparison with other schemes in the literature, as shown in Table \ref{TablaIC}. The value of $V(0)$ is obtained considering the log of the power of the signal $y_t$ (for further details, see the alternative BIC computation with Gaussian noise in \cite[page 375]{Priestley81}).
\newline
\newline
 {\bf Remark.} Note this example satisfies all the assumptions in the derivation of BIC. Therefore, we expect excellent results of BIC. We desire to test the performance of UAED, even in this scenario where BIC is clearly favored. 
\newline
\newline 
 The decided orders of the model given by AIC, BIC, HQIC, and UAED in the different scenarios and simulations are given from Figure \ref{ExARfig1} to Figure \ref{ExARfig12}. All these figures show the histograms of the decided orders given in each run by the different methods. Namely, the bars represent the percentages of times that an index is chosen as order of the model.
   Each figure corresponds to a specific scenario with a true order of the model $k^*\in\{3,5\}$, a certain number of data $T\in\{200, 2000\}$, and a noise level $\sigma_\epsilon \in\{0.5,1,2\}$. Table \ref{TablaSummaryRes} summarizes the results, showing the correct-decision rate $p_A\in[0,1]$, and using the following ranking for the performance:
 \begin{itemize}
\item {\bf Best:}  for the method that provides the highest  correct-decision rate $p_A$.
\item {\bf Excellent:}  for the methods with $p_A\geq0.95$.
\item {\bf Good:}  for the methods with  $ 0.80 \leq p_A < 0.95$.
\item {\bf Fair:}  for the methods with $ 0.50 \leq p_A < 0.80$.
\item {\bf Poor:}  for the methods with $0.20 \leq p_A< 0.50$.
\item {\bf Bad:}  for the methods with $0.10 \leq p_A< 0.20$.
\item {\bf Very bad:}  for the methods with $p_A< 0.10$.
\end{itemize}
We can observe that BIC and UAED provide the best performance. As remarked above, this example is particularly favorable for BIC but, even in this numerical experiment, UAED provides very good results and, in some scenarios, even  the best results. Indeed, as we can observe from Table \ref{TablaSummaryRes}, UAED gives the best results with $T=2000$ (i.e., when we have more data), regardless of the noise level and the true order of the model.
 It is important to highlight that well-known methods, such as AIC and HQIC, provide very poor performance. Clearly, additional information regarding the dispersion of the wrong decisions can be observed in  Figures  from \ref{ExARfig1} to \ref{ExARfig12}. This experiment shows clearly that UAED is a competitive and robust methodology.

\begin{table}[!h]	
	\caption{Summary of the results in the example in Section \ref{ARexample}.}\label{TablaSummaryRes}
	\vspace{-0.5cm}
	\footnotesize
	\begin{center}
		\begin{tabular}{|c|c|c||c|c|c|c|c|c|} 
			\hline 
			 \multicolumn{3}{|c||}{{\bf Scenario}} & \multicolumn{4}{c|}{{\bf Method}}   &   \\ 
			 \hline
			  $k$ & $\sigma_\epsilon$ & $T$ & {\bf UAED } & {\bf BIC} &  {\bf AIC } & {\bf HQIC} &  {\bf Fig.}   \\ 
			\hline 
			\hline 
	 		\multirow{6}{*}{$3$}  & \multirow{2}{*}{ $0.5$}  & $200$ & Good, $p_A\approx 0.82$   & \cellcolor{green!15}{\bf Best}, $p_A\approx 0.97$  &  Bad, $p_A\approx 0.13$  & Very bad, $p_A\approx 0.01$   & \ref{ExARfig1} \\
			 &  & $2000$ &    \cellcolor{green!15}{\bf Best}, $p_A= 1$ & \cellcolor{green!15} {\bf Best}, $p_A= 1$  &  Bad, $p_A\approx 0.1$  &  Bad, $p_A\approx 0.1$   &  \ref{ExARfig2} \\
			 \cline{2-8}
			 & \multirow{2}{*}{ $1$}   & $200$ &  Good, $p_A\approx 0.82$   &  \cellcolor{green!15} {\bf Best}, $p_A\approx 0.97$  &  Bad, $p_A\approx 0.13$  & Very bad, $p_A\approx 0.01$    &  \ref{ExARfig3} \\
			 &   & $2000$ &  \cellcolor{green!15} {\bf Best}, $p_A= 1$ & Excellent, $p_A\approx 0.98$  &  Bad, $p_A\approx 0.13$  &  Bad, $p_A\approx 0.17$  &  \ref{ExARfig4} \\
			  \cline{2-8}
			 & \multirow{2}{*}{ $2$} & $200$ &  Good, $p_A\approx 0.81$   &  \cellcolor{green!15} {\bf Best}, $p_A\approx 0.96$  &  Bad, $p_A\approx 0.11$  & Very bad, $p_A\approx 0.01$   &  \ref{ExARfig5} \\
			 &   & $2000$ &   \cellcolor{green!15}{\bf Best}, $p_A= 1$ & Excellent, $p_A\approx 0.97$  &  Very bad, $p_A\approx 0.05$  &  Very bad, $p_A\approx 0.06$  &  \ref{ExARfig6}\\
			\hline 
			\hline
			\multirow{6}{*}{$5$}  & \multirow{2}{*}{ $0.5$}  & $200$ &  Good, $p_A\approx 0.80$   &  \cellcolor{green!15} {\bf Best}, $p_A\approx 0.90$  &  Bad, $p_A\approx 0.14$  & Very bad, $p_A\approx 0.01$ &  \ref{ExARfig7}\\
			 &   & $2000$ &  \cellcolor{green!15}  {\bf Best}, $p_A= 1$ &  \cellcolor{green!15} {\bf Best}, $p_A= 1$  &  Very bad, $p_A\approx 0.07$  &  Bad, $p_A\approx 0.1$   &  \ref{ExARfig8}\\
			 \cline{2-8}
			 & \multirow{2}{*}{ $1$} & $200$ &   Good, $p_A\approx 0.80$   &  \cellcolor{green!15} {\bf Best}, $p_A\approx 0.90$  &  Bad, $p_A\approx 0.17$  & Very bad, $p_A\approx 0.03$   &  \ref{ExARfig9} \\
			&   & $2000$ &  \cellcolor{green!15} {\bf Best}, $p_A= 1$ & Excellent, $p_A\approx 0.98$  &  Bad, $p_A\approx 0.10$  &  Bad, $p_A\approx 0.13$   &  \ref{ExARfig10} \\
			\cline{2-8}
			 & \multirow{2}{*}{ $2$} & $200$ &   Good, $p_A\approx 0.80$   &  \cellcolor{green!15} {\bf Best}, $p_A\approx 0.90$  &  Bad, $p_A\approx 0.13$  & Very bad, $p_A\approx 0.01$  &  \ref{ExARfig11} \\
			 &  & $2000$ &  \cellcolor{green!15}{\bf Best}, $p_A= 1$ & Excellent, $p_A\approx 0.98$  &  Very bad, $p_A\approx 0.09$  &  Bad, $p_A\approx 0.11$  &  \ref{ExARfig12}\\
			\hline
		\end{tabular}	
	\end{center}
\end{table}

\subsection{Variable selection in a regression problem with real data}\label{VSwithRealData}

In this section, we present a feature selection problem for regression. Moreover, we consider real data.  
 More specifically, a dataset of $N$ pairs  $\{{\bf x}_n,y_n\}_{n=1}^N$ is given, where each input vector ${\bf x}_n=[x_{n,1},...,x_{n,K}]$ is formed by $K$ variables, and the outputs  $y_n$'s are scalar values. We assume $K\leq N$ and a linear observation model,
\begin{align}\label{aquiM0}
    y_n&=\theta_0+\theta_1 x_{n,1}+\theta_2 x_{n,2}+...\theta_K x_{n,K}+ \epsilon_n,  
\end{align}
 where $ \epsilon_n$ is Gaussian noise with zero mean and variance $\sigma_\epsilon^2$, i.e.,  $ \epsilon_n \sim \mathcal{N}(\epsilon|0,\sigma_\epsilon^2)$. In the real dataset studied in \cite{OurPaperSound}, there are $K=122$ features and $N=1214$ number of data points. The output represents the variable defined as ``arousal'' in \cite{OurPaperSound}.
\newline
 In order to allow the comparison with other schemes in the literature, here we can set  $V(k)=-2\log(\ell_{\texttt{max}})$ where $\ell_{\texttt{max}}=\max_{\bm \theta} p({\bf y}| {\bm \theta}_k)$ with $k\leq K$, after ranking the 122 variables as in \cite{OurPaperSound}. Clearly, The likelihood function $p({\bf y}| {\bm \theta}_k)$ is induced by Eq. \eqref{aquiM0}.
Therefore, in this experiment, we can compare UAED again with other information criterion measures in the literature, some of them are given in Table \ref{TablaIC}. 
BIC suggests a model with 17 variables, AIC chooses 44 variables, and HQIC selects 41 variables. The proposed UAED suggests considering only 11 variables. Therefore, the UAED suggestions is closer to the results given in other previous studies and to experts' recommendations in the literature, e.g., \cite{OurPaperSound}.

\subsection{Variable selection in a classification problem with a nonalcoholic fatty liver disease real dataset}\label{VSwithOscar}

\begin{figure}[h!]
\centerline{
\subfigure[\label{orderPolLabel}]{\includegraphics[width=9cm]{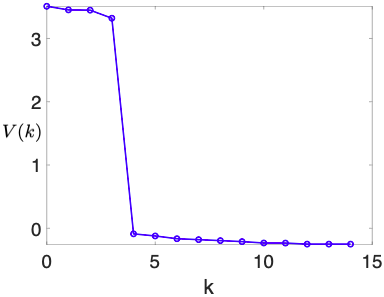}}
\subfigure[\label{OscarLabel}]{\includegraphics[width=9cm]{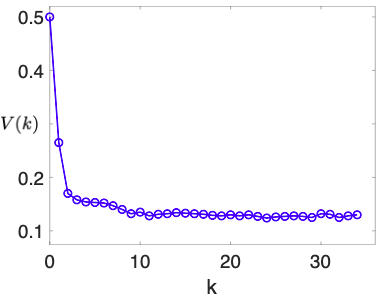}}
}
\caption{ {\bf (a)} The corresponding curve $V(k)=-2\log \ell_{\texttt{max}}$ (with $\ell_{\texttt{max}}=\max_{\bm{\theta}} p({\bf y}| \bm{\theta}_k)$) in Section \ref{PolOrder};  {\bf (b)} The curve $V(k)=1-\mbox{accuracy}(k)$ of the experiment in Section \ref{VSwithOscar}. 
}
\label{Oscar_app}
\end{figure}

Biomedical applications are nowadays extremely relevant \cite{Ali22,Laghari22}.
The authors in \cite{OscarPaper} analyze the most important features for predicting patients at risk
of developing nonalcoholic fatty liver disease. The authors collected data from 1525 patients who attended the Cardiovascular Risk Unit of Mostoles University Hospital (Madrid, Spain) from 2005 to 2021, and use a random forest (RF) algorithm to classify patients and rank the input features, in order to select the most important one. They found that $4$ features were the most relevant according to the ranking and the experts' opinions: (a) insulin resistance, (b) ferritin, (c) serum levels of insulin, and (d) triglycerides. 
 \newline
In this experiment, we set  $V(k)=1-\mbox{accuracy}(k)$ that is given in Figure \ref{OscarLabel}, after ranking the 35 features \cite{OscarPaper}. Note  that $V(0)=0.5$ representing a completely random binary classification. It is important to remark that, with this choice of $V(k)$, the other information criteria cannot be employed. The application of UAED suggests to select $4$ variables which is exactly the result of the paper \cite{OscarPaper}, obtained using a cross-validation approach, and supported by the experts' opinions.


\subsection{Variable selection in a real dataset of ventricular fibrillation}\label{VS_BIO_2}
The authors in \cite{EduBioVS2016} addressed the early recognition of ventricular fibrillation (VF). In their study, they considered a feature selection based on the percentage of the balanced error rate (BER), which plays the role of $V(k)$. See Equation 5 at page 7 of \cite{EduBioVS2016} and Figure 4-(a)-right (green curve) at page 10 of \cite{EduBioVS2016} for further information regarding this $V(k)$ curve. We consider as $V(0)$ the first point provided in this curve.
\newline
 It is important to remark that again, with this choice of $V(k)$ based on the balanced error rate, the other information criteria cannot be employed. The application of UAED suggests to choose $6$ features which is very close to the suggestion of the authors in \cite{EduBioVS2016} that was $5$ features. The authors in \cite{EduBioVS2016} reached this conclusion after using much more complex analyses and comparing with expert's opinions. In this sense, UAED is also a simpler approach to employ.
 

\section{Conclusions}\label{sec:conclusions}

A novel automatic elbow detector for model selection purposes has been introduced. The proposed UAED scheme is inspired by the concept of the maximum ``area under the curve'' (AUC) in receiver operator characteristic (ROC) curves. The contributions of the work can be divided into four main parts: (a) motivation and derivations of the proposed method, (b) analysis of the behavior in ideal scenarios and its properties, (c) test and comparison with several numerical simulations, and (d) a related Matlab code.
\newline
Four different geometrical derivations of UAED have been provided. The first three derivations are based on the vertical, horizontal and Euclidean distance of the $V(k)$ curve (with $k \in \mathbb{N}$) with respect to a decreasing straight line, which corresponds to the ideal decay when each component has equal importance. The last derivation shows an additional property of the proposed solution, in a generalized framework with $k$ is a continuous variable instead of an integer (as in the rest of the work). This property can play a relevant role in future works regarding this research line. 
\newline
Furthermore, we have analyzed other features and properties of UAED, as the invariance on scaling the axes and the behavior in ideal scenarios.  The relationships and differences with several information criteria (already given in the literature) have been described and highlighted. More specifically, UAED can be also considered as an information criterion with the choice of the slope of the model penalty as $\lambda=\frac{V(0)}{k_{\texttt{max}}}$. However, the $V(k)$ curve in UAED can be  also chosen differently from the usual definition
$V(k)=-2\log \ell_{\texttt{max}}$ with  $\ell_{\texttt{max}}=\max_{\bm{\theta}} p({\bf y}| \bm{\theta}_k)$,
which is required in the other information criteria.  Hence, it is important to remark that the proposed procedure has a much wider range of application with respect to the other schemes in the literature, as also clarified by the numerical experiments. 
\newline 
Six experiments and comparisons show the benefits of the proposed UAED scheme. We have considered different examples in clustering, order selection, and variable selection within regression and classification problems. Moreover, three of these six experiments involve real datasets. We have compared with the most relevant information criteria in the literature (AIC, BIC and HQIC). Even in scenarios that are much favorable for one of them  (i.e., AIC, BIC or HQIC), the proposed UAED method provides very competitive results.



{\small
\section*{{\small Acknowledgement}}

The work was partially supported by the Young Researchers R\&D Project,  ref. num. F861 (AUTO-BA-GRAPH) funded by Community of Madrid and Rey Juan Carlos University, and by Agencia Estatal de Investigaci{\'o}n AEI (project SP-GRAPH, ref. num. PID2019-105032GB-I00).
}

%
%

\bibliographystyle{IEEEtran}

\bibliography{bibliografia.bib}

\appendix
\section{Third alternative derivation}\label{GrandeLucaEdu}
Let us consider Figure \ref{Fig4Luca}. First of all, we must find the value $k'$ such that the straight line, connecting the points $(0,V(0))$ and $(k_{\texttt{max}},0)$, which reaches the value $V(k)$ (where $k\neq k'$, and more precisely $k\leq k'$). Namely, we desire to obtain $k'$ such that
 \begin{align}
 V(k)=-\frac{V(0)}{k_{\texttt{max}}}\cdot k'+V(0),
 \end{align}
 hence
 \begin{align}
 k'=-\frac{k_{\texttt{max}}}{V(0)}\left[V(k)-V(0)\right].
 \end{align}
Now, we could also consider to maximize the following difference
 \begin{align}
 r(k)&=k'-k =-\frac{k_{\texttt{max}}}{V(0)}\left[V(k)-V(0)\right]-k,
 \end{align}
 and the elbow is defined as
 \begin{align}
k^*=\arg\max_k \  r(k)&=\arg\max_k\left[ -\frac{k_{\texttt{max}}}{V(0)}V(k)-k\right], \nonumber \\
&=\arg\min_k\left[ \frac{k_{\texttt{max}}}{V(0)}V(k)+k\right], \nonumber \\
&=\arg\min_k\left[ V(k)+\frac{ V(0)}{k_{\texttt{max}}}k\right], \label{MagicLucaEdu}
 \end{align}
where in the last we have multiplied by the constant $V(0)$. Note that Eq. \eqref{MagicLucaEdu} is exactly the same optimization problem (i.e., with the same cost function) in Sections \ref{FirstDerSect}-\ref{GrandeEdu}.

\section{Fourth alternative derivation}\label{otherder}
\begin{figure}[!ht]
   \centerline{ 
   \subfigure[\label{Fig4Luca}]{\includegraphics[width=0.5\columnwidth]{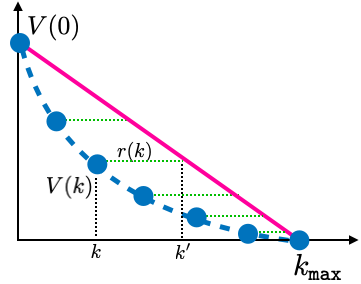}}
   
   \subfigure[\label{figLucaMuchas_2}]{\includegraphics[width=0.5\columnwidth]{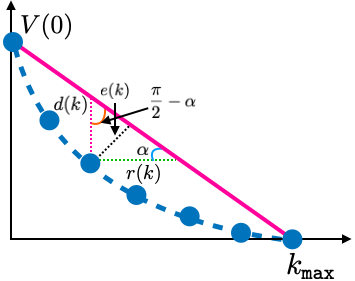}}          
  }
   \caption{{\bf (a)} Graphical representation of the other alternative derivation in Appendix \ref{GrandeLucaEdu}. {(\bf b)} Graphical representation of  derivation based on the Euclidean distance $e(k)$.}
   \label{figLucaMuchas_2_all}
\end{figure}

One could also consider the Euclidean distance $e(k)$ between the points in the curve $V(k)$ and the straight line connecting the points $(0,V(0))$ and $(k_{\texttt{max}},0)$, as depicted in Figure \ref{figLucaMuchas_2}.
Observing this figure, we can notice that 
\begin{align}
e(k)&=d(k) \sin(\pi/2-\alpha)=d(k) \cos\alpha =r(k) \sin\alpha,
\end{align}
where $\alpha$ is the angle shown in Figure \ref{figLucaMuchas_2}.
Since the angle $\alpha$ is constant, then we can write 
\begin{align}
k^*=\arg \max_k \left[e(k)\right]&=\arg \max_k \left[d(k) \cos\alpha \right]=\arg \max_k \left[d(k)\right], \nonumber \\
&=\arg \max_k \left[r(k) \sin\alpha\right] =\arg \max_k \left[r(k)\right].
\end{align}
Therefore, maximizing $e(k)$ is equivalent to maximize $d(k)$ or $r(k)$. 
 
\section{An additional property}\label{AppSuperLuca}

\begin{figure}[!h]
   \centerline{ 
    \subfigure[\label{figLucaSuperFIGa}]{\includegraphics[width=0.5\columnwidth]{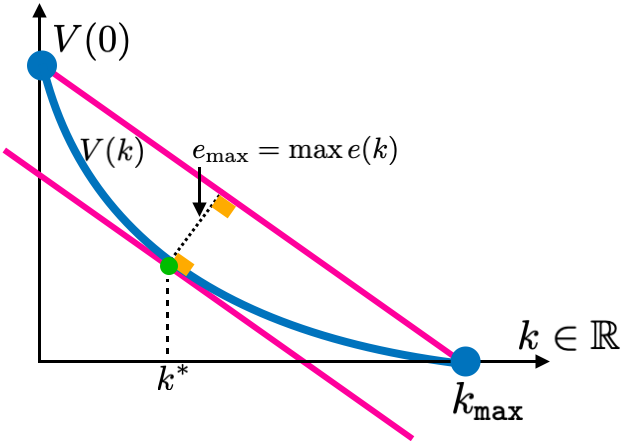} } 
    \subfigure[\label{figLucaSuperFIGb}]{\includegraphics[width=0.5\columnwidth]{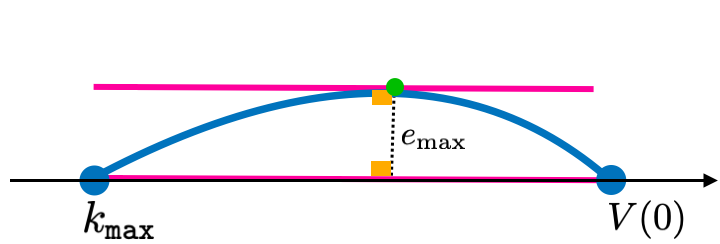}}          
  }
   \caption{An additional property of UAED, when we consider $k$ as a continuous parameter, i.e., $k\in \mathbb{R}$, instead of a discrete variable. The derivative ${\dot V}(k)=\frac{d V}{d k}$ evaluated at the optimal value $k^*$ (obtained by UAED) is equal to the slope of the straight line passing through the points $(0, V(0))$ and $(k_{\texttt{max}},0)$, i.e., ${\dot V}(k^*)=-\frac{V(0)}{k_{\texttt{max}}}$.}
   \label{figLucaSuperFIG}
\end{figure}

So far, we have considered that $k$ is a discrete variable. If we assume that $k$ can be a continuous parameter, i.e., $k\in \mathbb{R}$, we can obtain an additional property of the UAED solution. For the sake of simplicity, we also assume that $V(k)$ is convex. This condition can be relaxed, and the following discussion can be generalized for more general curves $V(k)$.
\newline
With these assumptions, it is possible to show that  the derivative ${\dot V}(k)=\frac{d V}{d k}$ evaluated at the optimal value $k^*$ (namely, the solution obtained by UAED) is equal to the slope of the straight line passing through the points $(0, V(0))$ and $(k_{\texttt{max}},0)$, i.e., 
$$
{\dot V}(k^*)=-\frac{V(0)}{k_{\texttt{max}}}.
$$
 Indeed, we know from App. \ref{otherder} that  $k^*=\arg \max_k \left[e(k)\right]$ where $e(k)$ is the Euclidean distance between the points $(k,V(k))$ and the straight line passing through $(0, V(0))$ and $(k_{\texttt{max}},0)$, as shown again in Figure \ref{figLucaSuperFIGa}. Applying a rotation to the plot in Figure \ref{figLucaSuperFIGa} and obtaining a new horizontal axis such that it coincides with  the straight line passing through $(0, V(0))$ and $(k_{\texttt{max}},0)$, we can observe that the optimal point must be a stationary point (i.e., with null derivative) in this new coordinate system (by construction). This is depicted in Figure \ref{figLucaSuperFIGb}. Therefore, the tangent straight line at the green point in Figure \ref{figLucaSuperFIGb} is parallel to the new horizontal axis. Inverting the rotation (namely, coming back the $k$-axis of Figure \ref{figLucaSuperFIGa}), the previous consideration is equivalent to say that the derivative ${\dot V}(k)=\frac{d V}{d k}$ evaluated at the optimal value $k^*$ must be  ${\dot V}(k^*)=-\frac{V(0)}{k_{\texttt{max}}}$, i.e., the tangent straight line at the green point is parallel to the straight line passing through the points $(0, V(0))$ and $(k_{\texttt{max}},0)$.

\section{Possible extension}
We have already shown that the resulting expression in Eq. \eqref{exp:N_opt} provides good performance and is endowed with valuable behaviors.
\newline
However, we can add more flexibility that can be useful in the scenarios in which the researchers and/or practitioners determine that the benefit of reducing the error is greater than the benefit of reducing the number of considered variables or vice versa. We define an additional parameter $\alpha \in [0,1]$, and consider the modified definition of the optimal $k$ as
\begin{equation}\label{exp:N_opt_general}
k^{*} = \arg\min_k\left[\alpha \cdot \frac{V(k)}{V(0)} + (1 - \alpha) \cdot \frac{k}{k_{\texttt{max}}} \right].
\end{equation}
Note that $\alpha= 0$ implies that all priority is to reduce the number of considered variables ($k^*=0$), that $\alpha=1$ implies that all priority is to reduce the resulting error (so that $k^*=k_{\texttt{max}}$). For   $\alpha = 0.5$, we come back to the definition in Eq. (\ref{exp:N_opt_2}). As we have previously done in Section \ref{sub:Calculo}, can rewrite Eq. \eqref{exp:N_opt_general} as
\begin{align}\label{exp:N_opt_general2}
k^{*} &=\arg\min_k\left[ V(k) +\underbrace{ \left(\frac{1 - \alpha}{\alpha}\frac{V(0)}{k_{\texttt{max}}}\right)}_{\lambda} \cdot k \right], \nonumber \\
&= \arg\min_k\left[ V(k) +\lambda  k \right], 
\end{align}
 having the form of an information criterion with a different choice of $\lambda$  which involves now the parameter $\alpha$, as well.

\begin{figure}[h!]
\centerline{
\subfigure[]{\includegraphics[width=5cm]{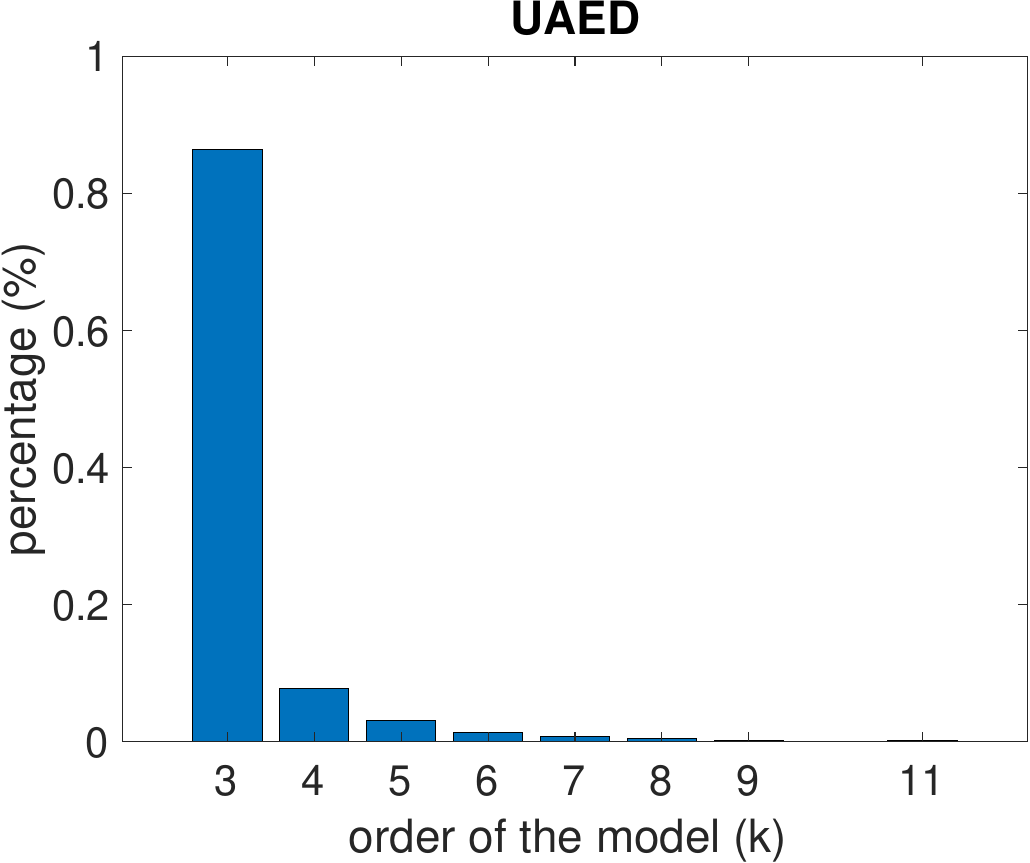}}
\subfigure[]{\includegraphics[width=5cm]{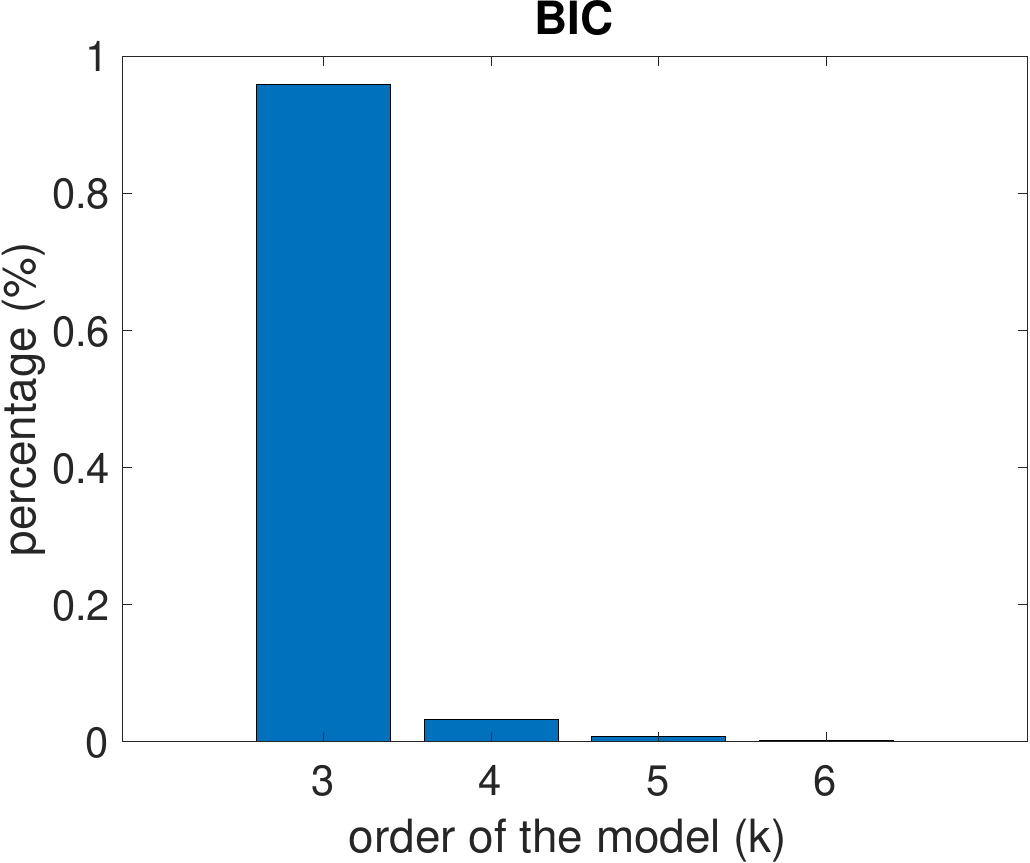}}
\subfigure[]{\includegraphics[width=5.1cm]{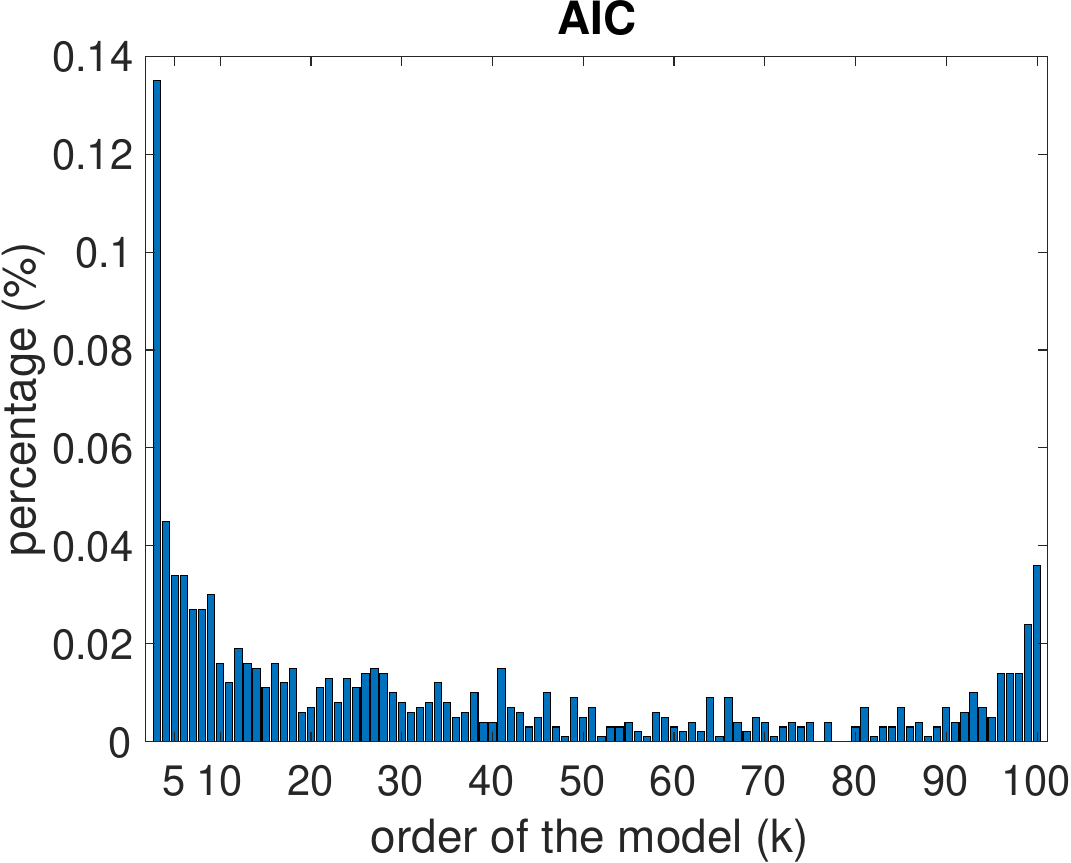}}
\subfigure[]{\includegraphics[width=5.1cm]{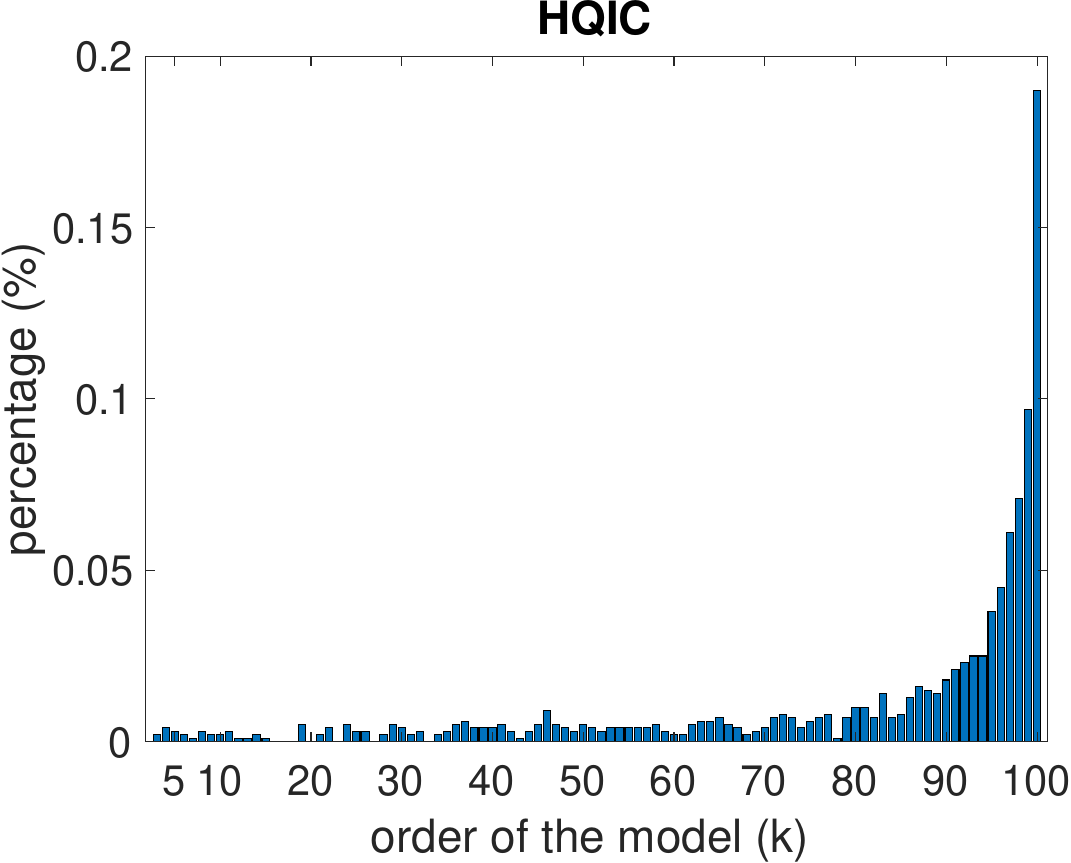}}
}
\caption{Percentages of model order decision by the different methods, in the scenario where the  true order is $k=3$,  the standard deviation of the noise is $\sigma_\epsilon=0.5$, and the number of data $T=200$; {\bf (a)} results of UAED; {\bf (b)} results of  BIC; {\bf (c)} results of  AIC;  {\bf (d)} results of HQIC.}
\label{ExARfig1}
\end{figure}

\begin{figure}[h!]
\centerline{
\subfigure[]{\includegraphics[width=5cm]{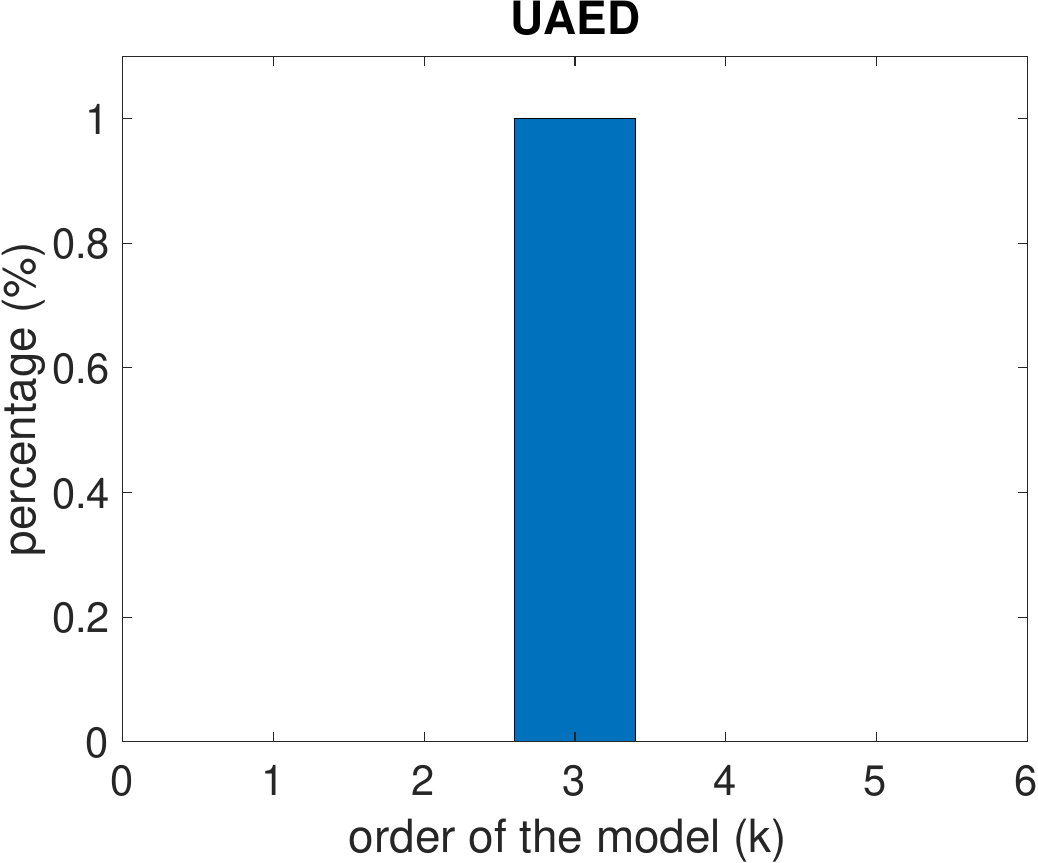}}
\subfigure[]{\includegraphics[width=5cm]{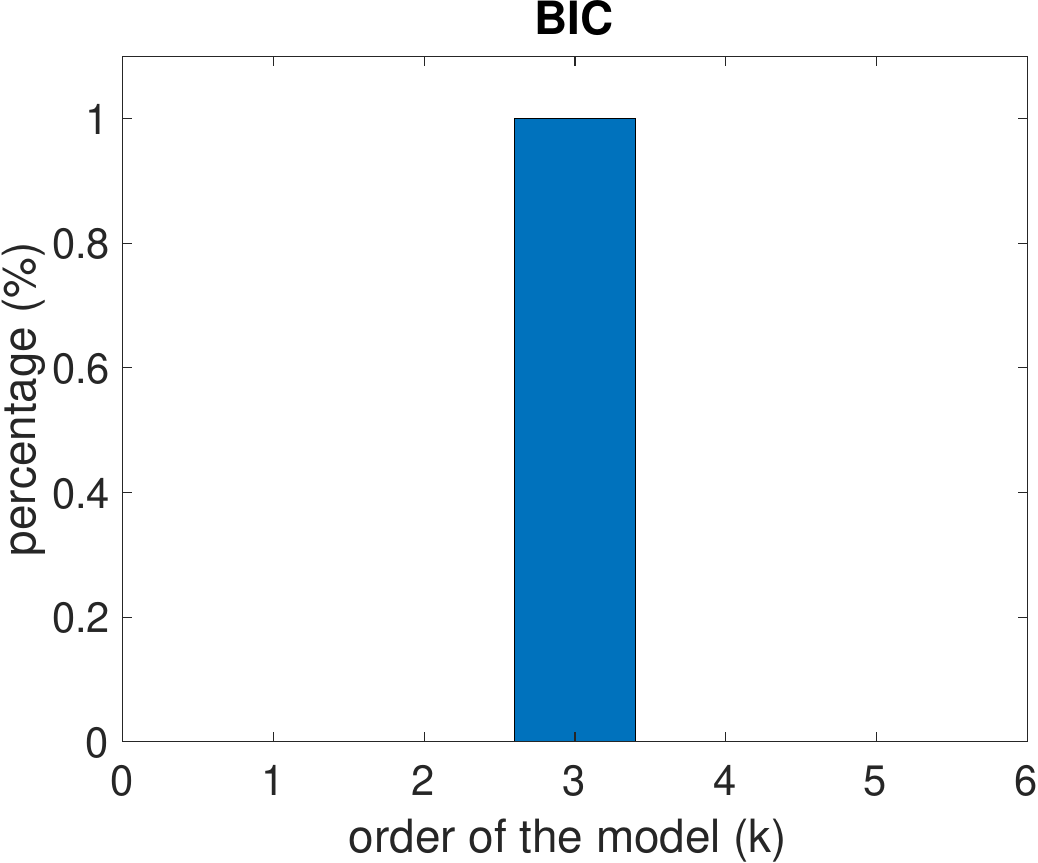}}
\subfigure[]{\includegraphics[width=5.1cm]{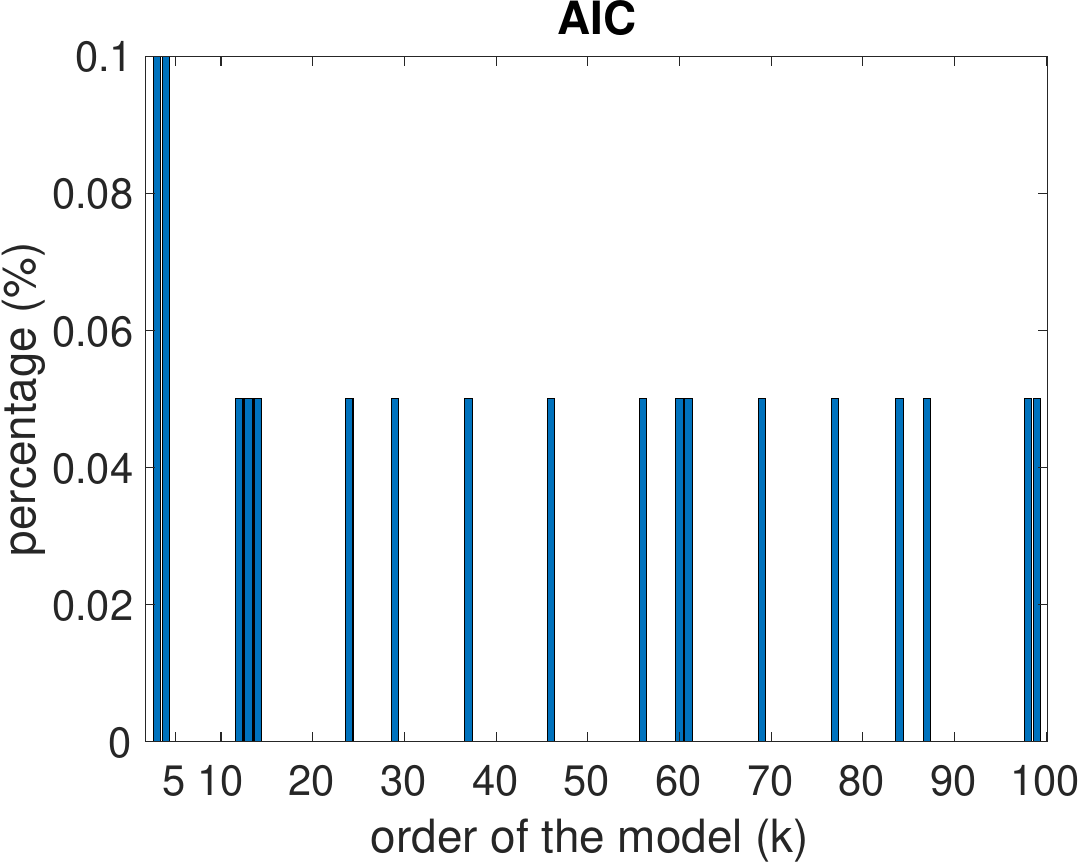}}
\subfigure[]{\includegraphics[width=5.1cm]{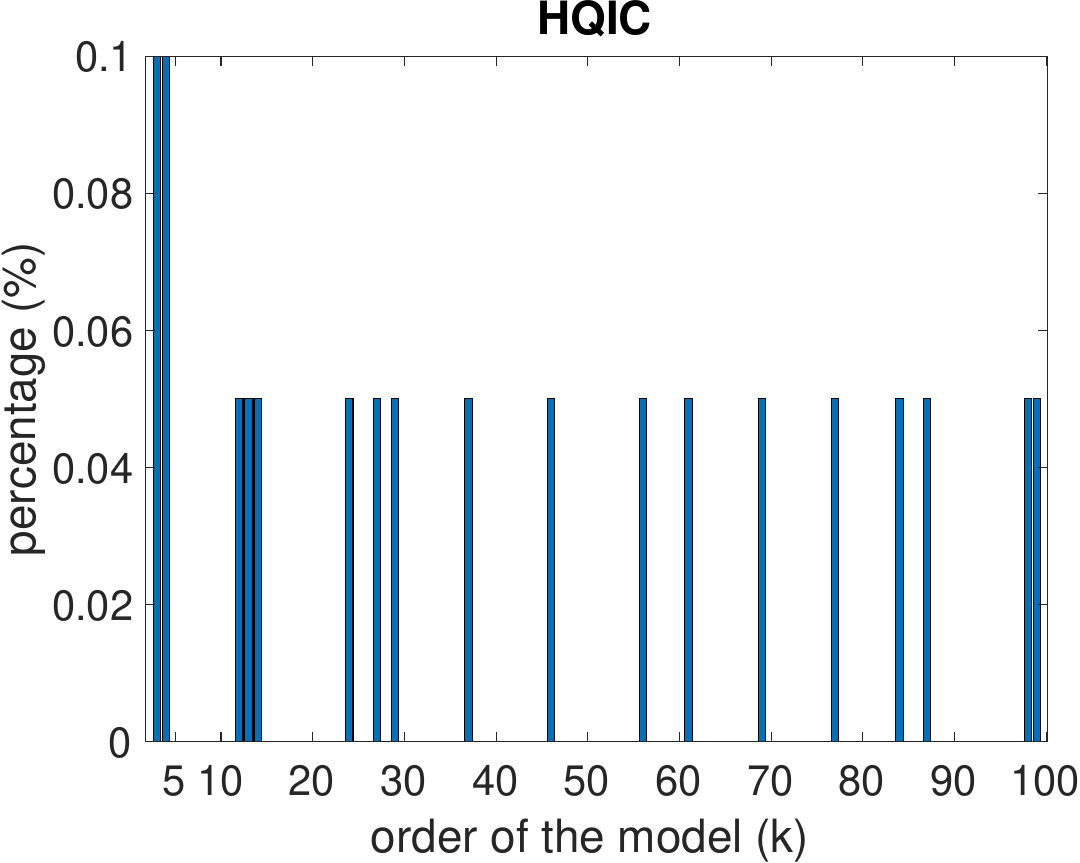}}
}
\caption{Percentages of model order decision by the different methods, in the scenario where the true order is $k=3$,  the standard deviation of the noise is $\sigma_\epsilon=0.5$, and the number of data $T=2000$; {\bf (a)} results of UAED; {\bf (b)} results of  BIC; {\bf (c)} results of  AIC;  {\bf (d)} results of HQIC.}
\label{ExARfig2}
\end{figure}

\begin{figure}[h!]
\centerline{
\subfigure[]{\includegraphics[width=5cm]{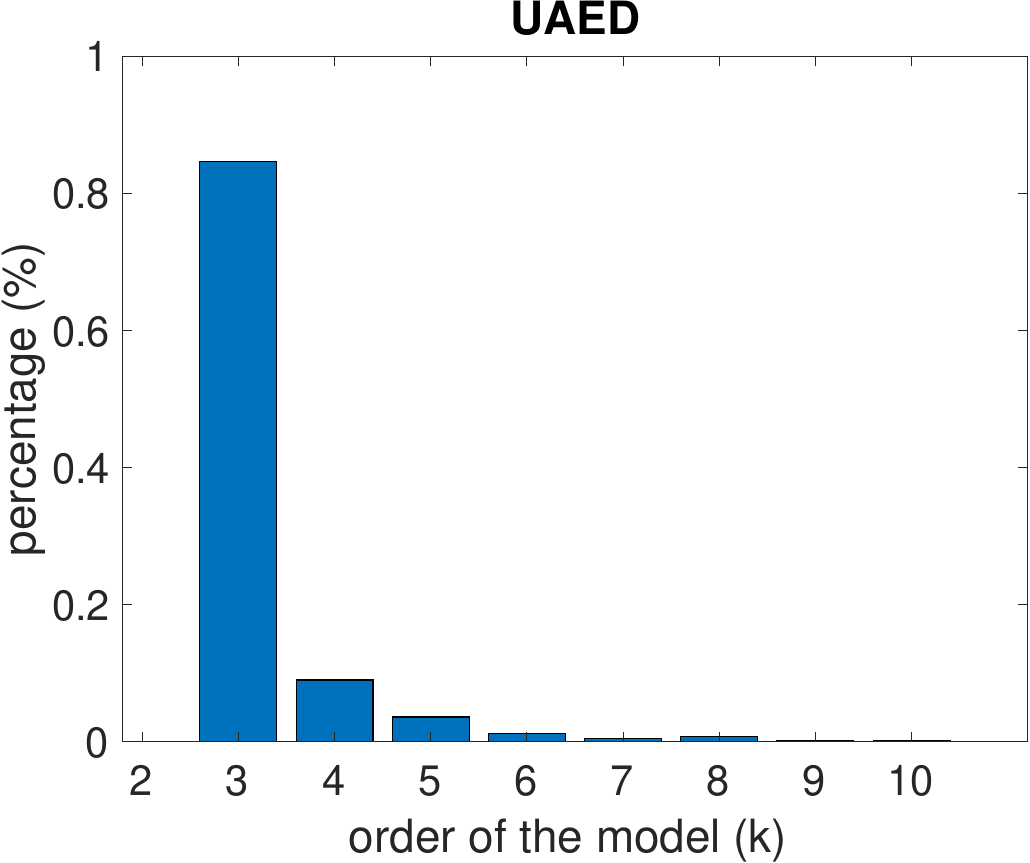}}
\subfigure[]{\includegraphics[width=5cm]{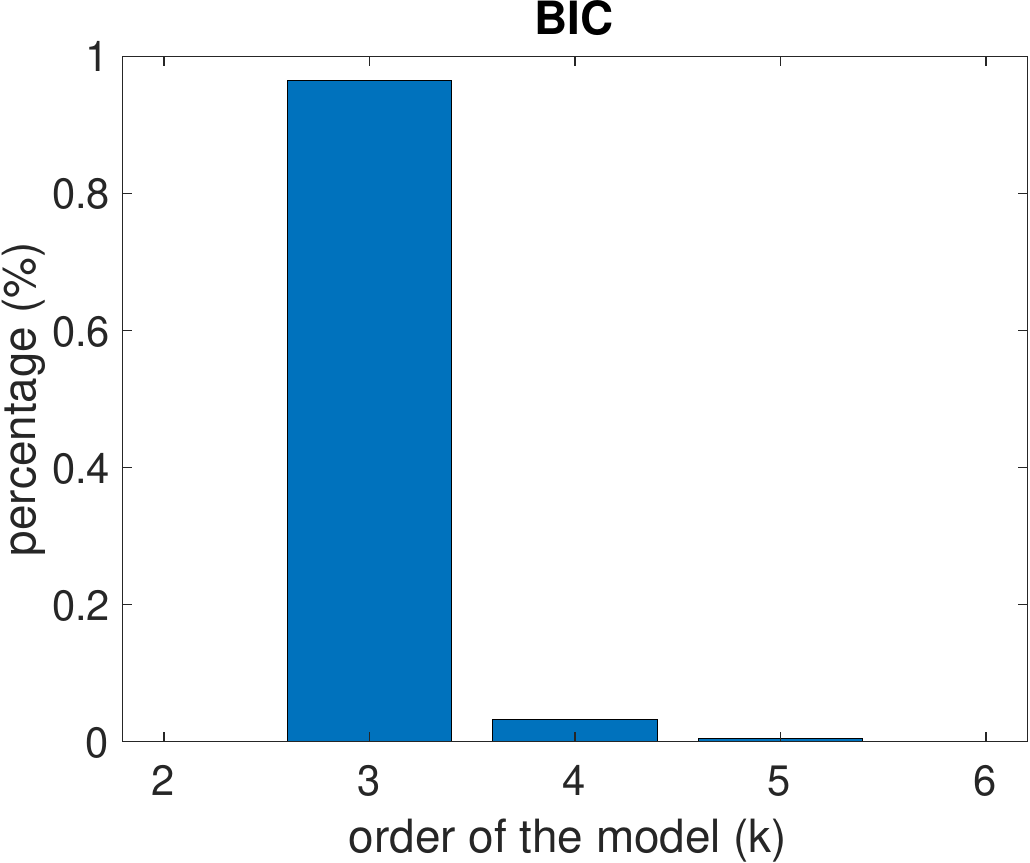}}
\subfigure[]{\includegraphics[width=5.1cm]{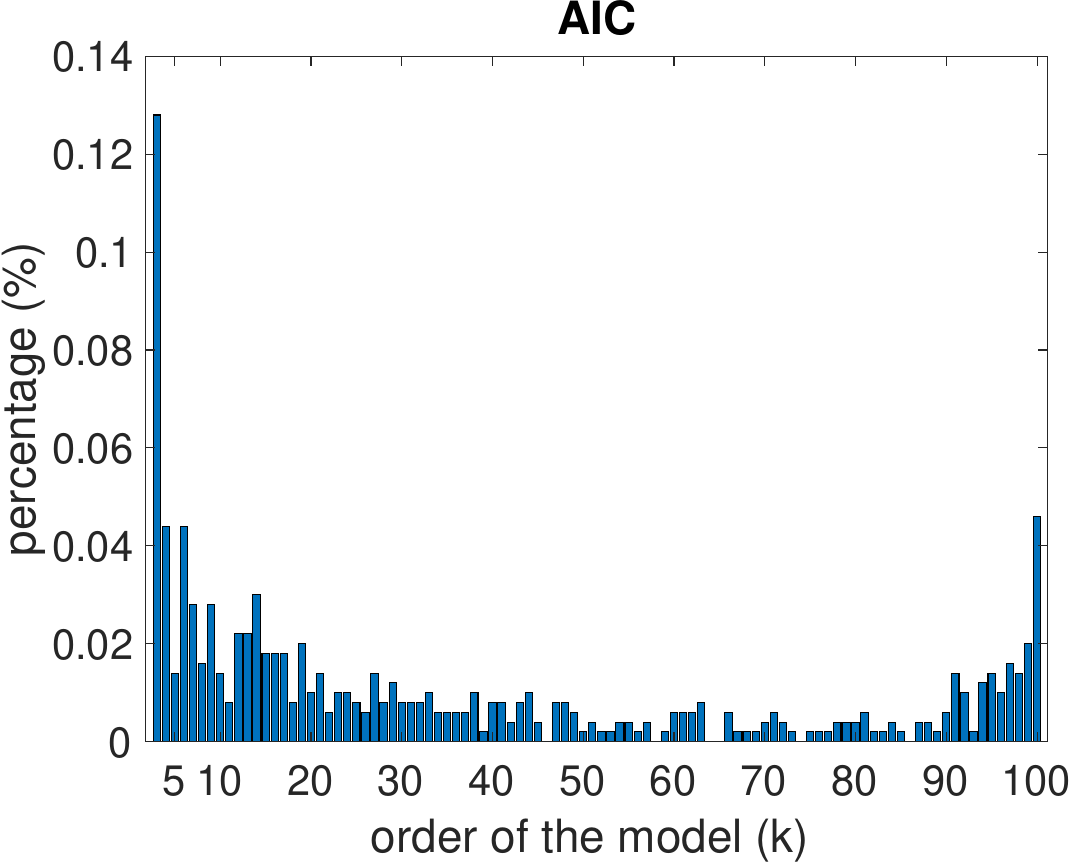}}
\subfigure[]{\includegraphics[width=5.1cm]{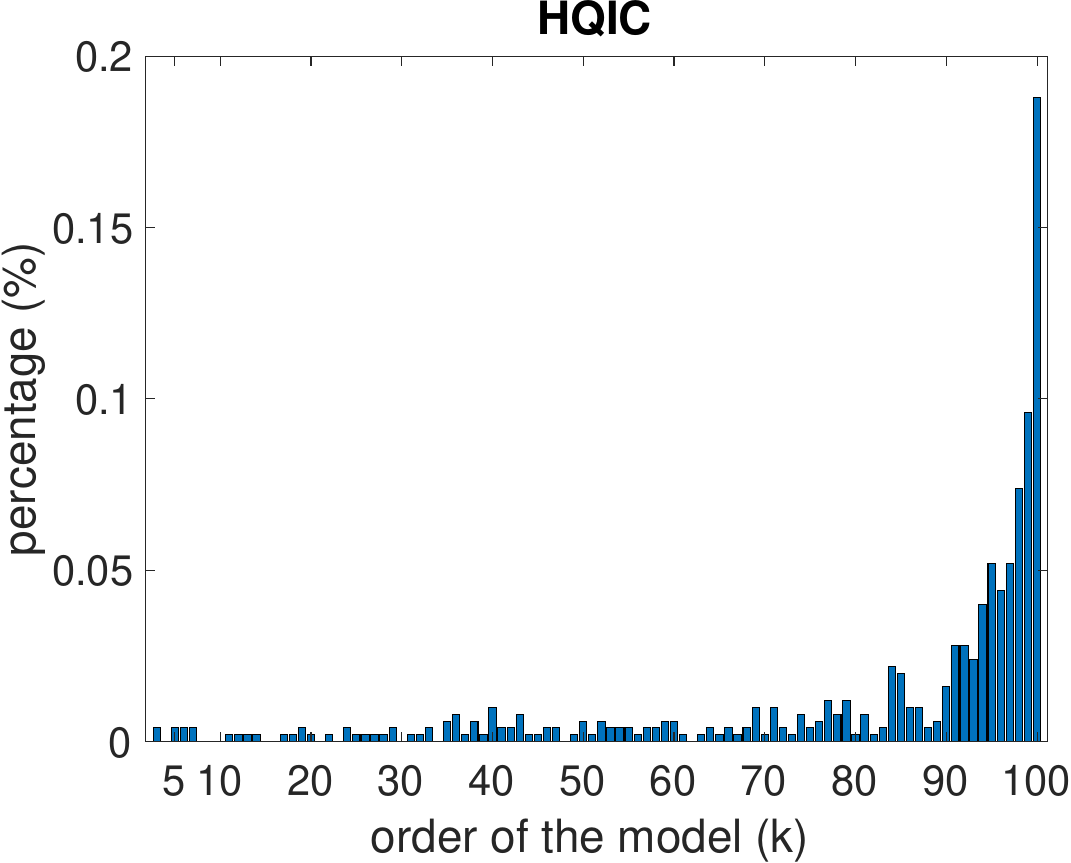}}
}
\caption{Percentages of model order decision by the different methods, in the scenario where the true order is $k=3$,  the standard deviation of the noise is $\sigma_\epsilon=1$, and the number of data $T=200$; {\bf (a)} results of UAED; {\bf (b)} results of  BIC; {\bf (c)} results of  AIC;  {\bf (d)} results of HQIC.}
\label{ExARfig3}
\end{figure}
\begin{figure}[h!]
\centerline{
\subfigure[]{\includegraphics[width=5cm]{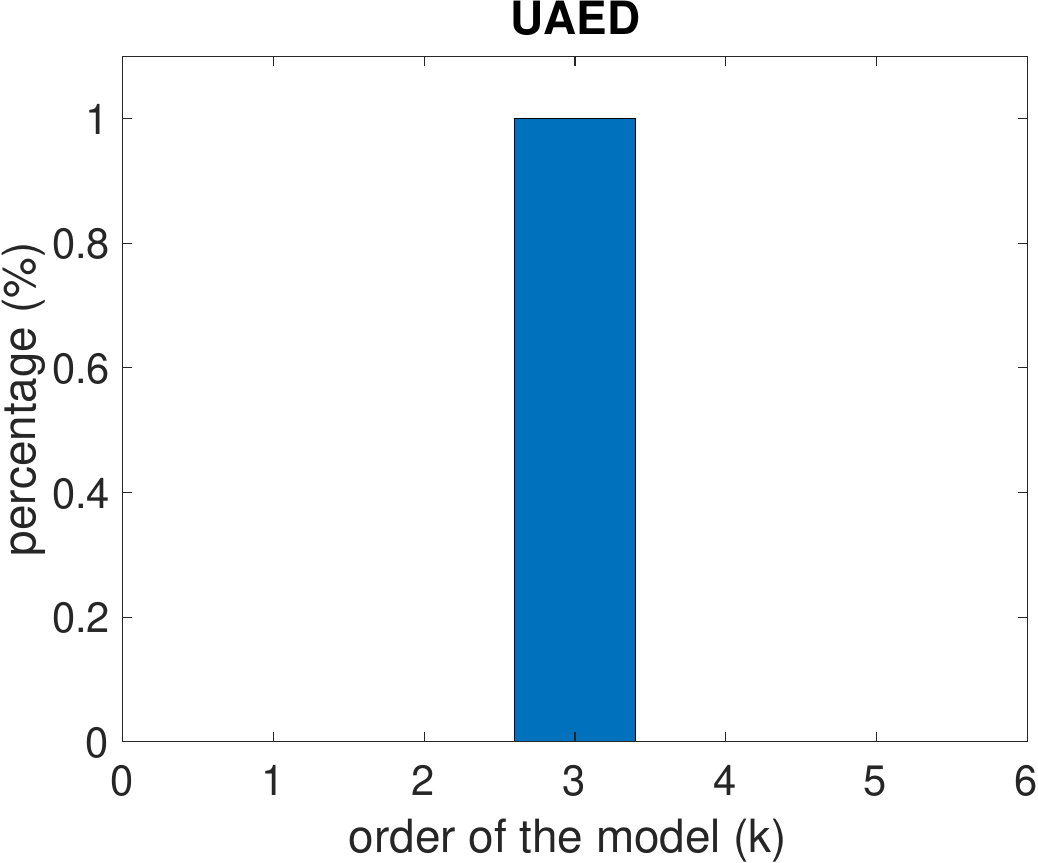}}
\subfigure[]{\includegraphics[width=5cm]{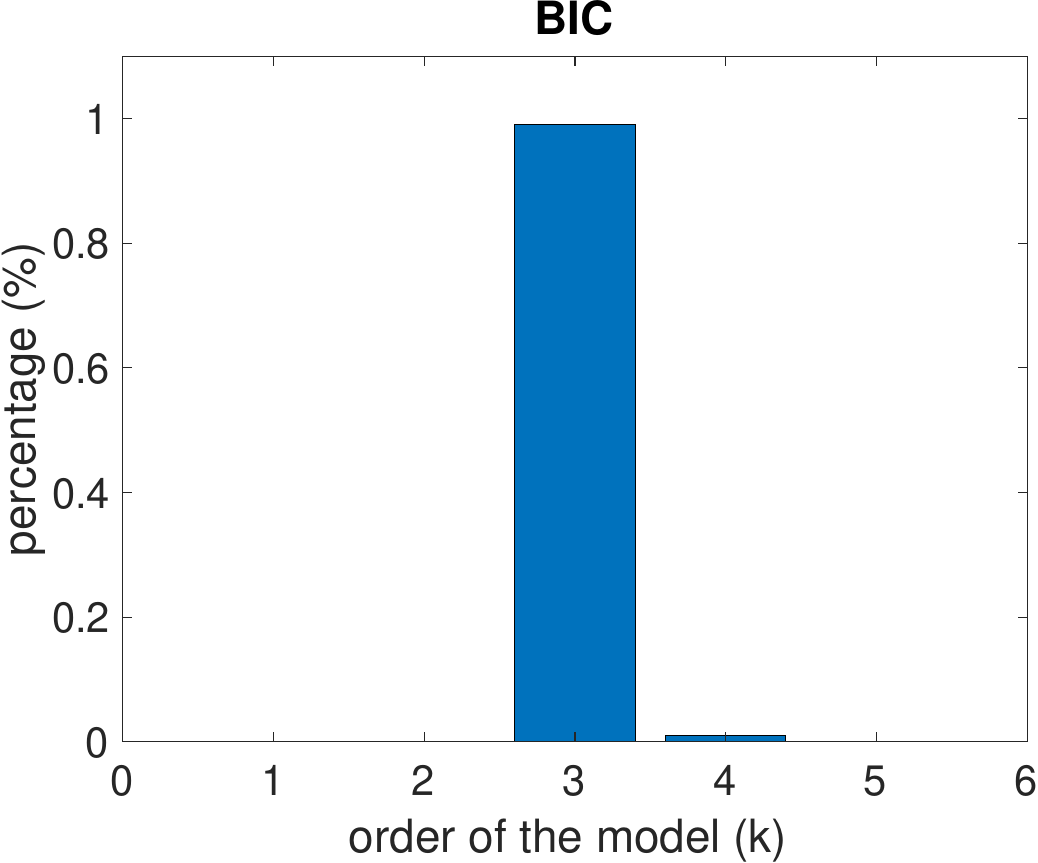}}
\subfigure[]{\includegraphics[width=5.1cm]{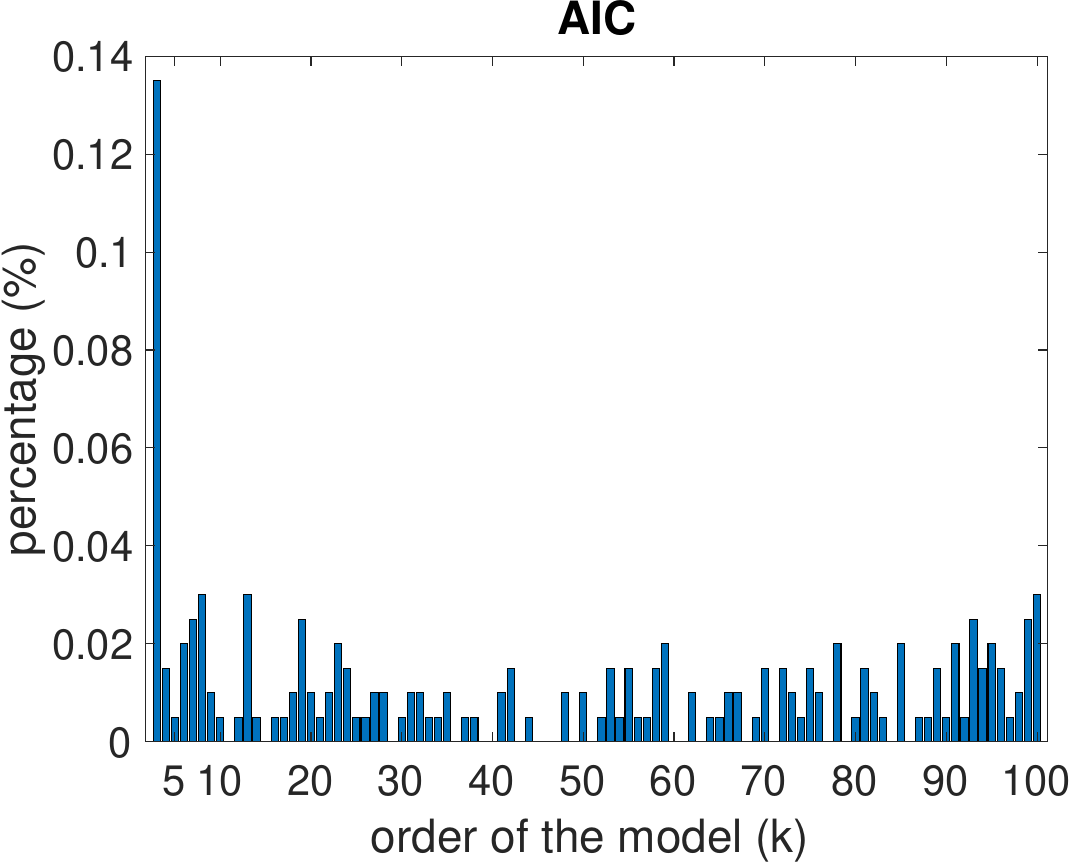}}
\subfigure[]{\includegraphics[width=5.1cm]{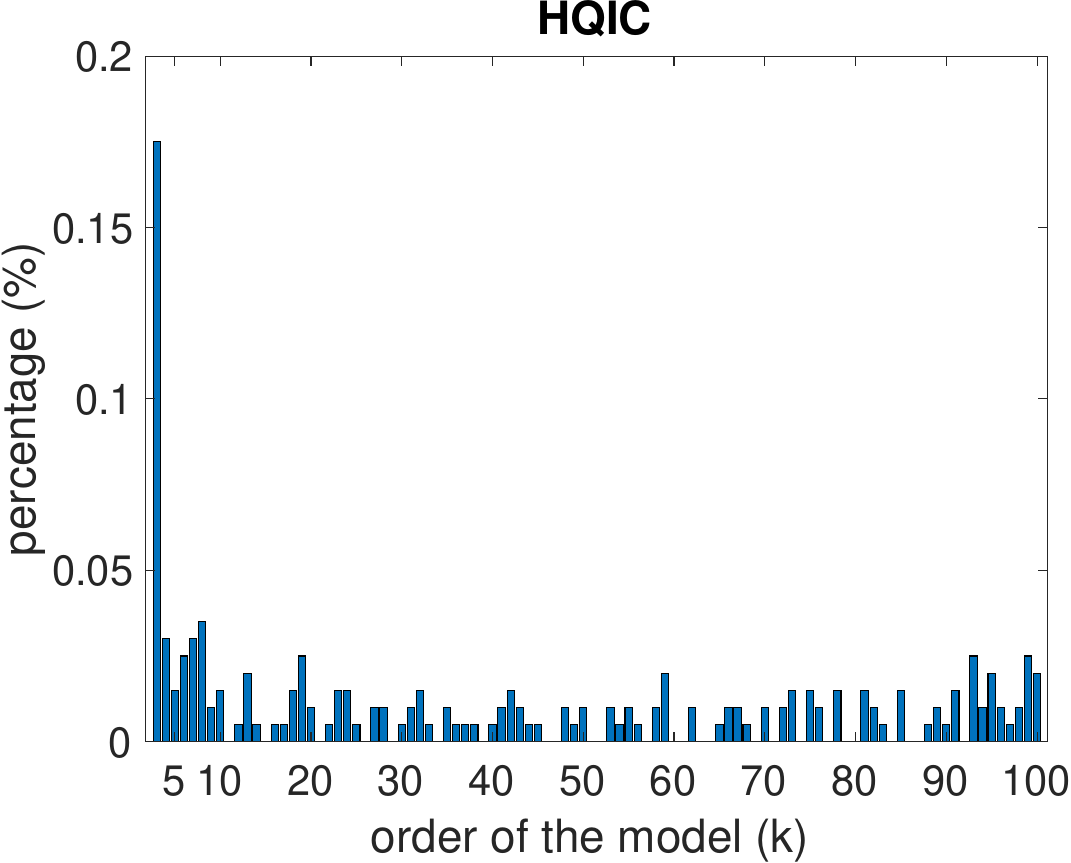}}
}
\caption{Percentages of model order decision by the different methods, in the scenario where the true order is $k=3$,  the standard deviation of the noise is $\sigma_\epsilon=1$, and the number of data $T=2000$; {\bf (a)} results of UAED; {\bf (b)} results of  BIC; {\bf (c)} results of  AIC;  {\bf (d)} results of HQIC.}
\label{ExARfig4}
\end{figure}
\begin{figure}[h!]
\centerline{
\subfigure[]{\includegraphics[width=5cm]{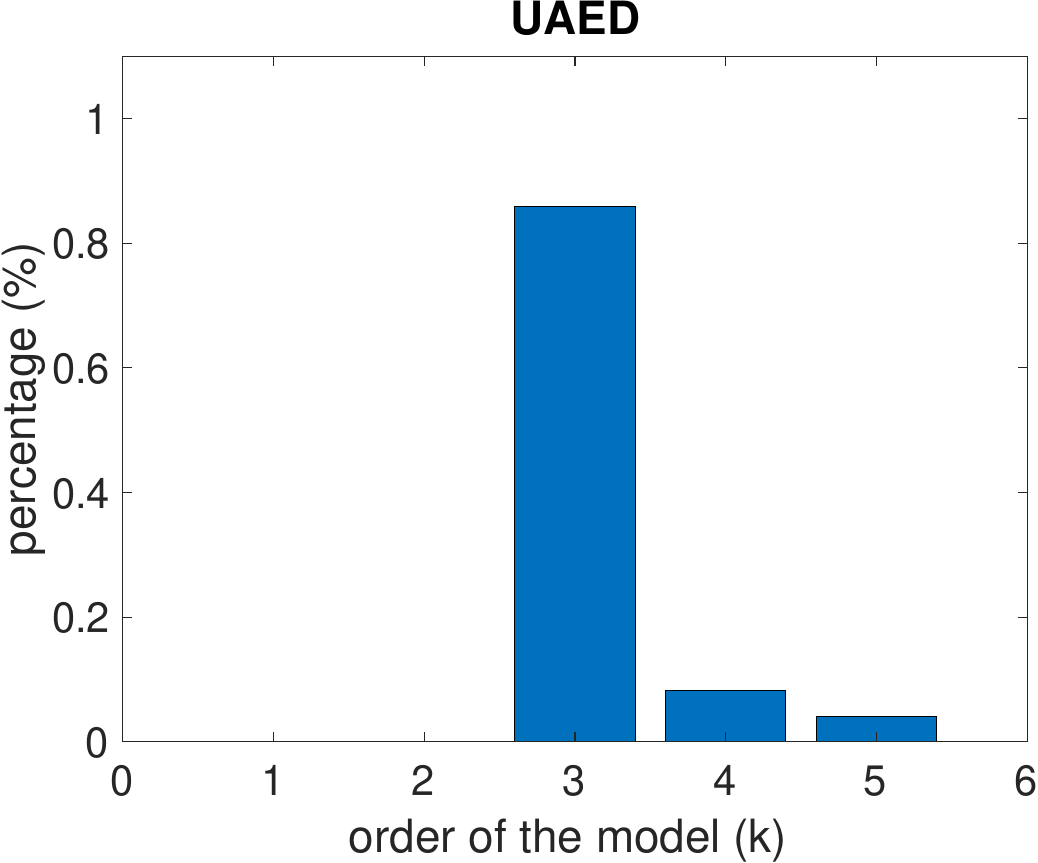}}
\subfigure[]{\includegraphics[width=5cm]{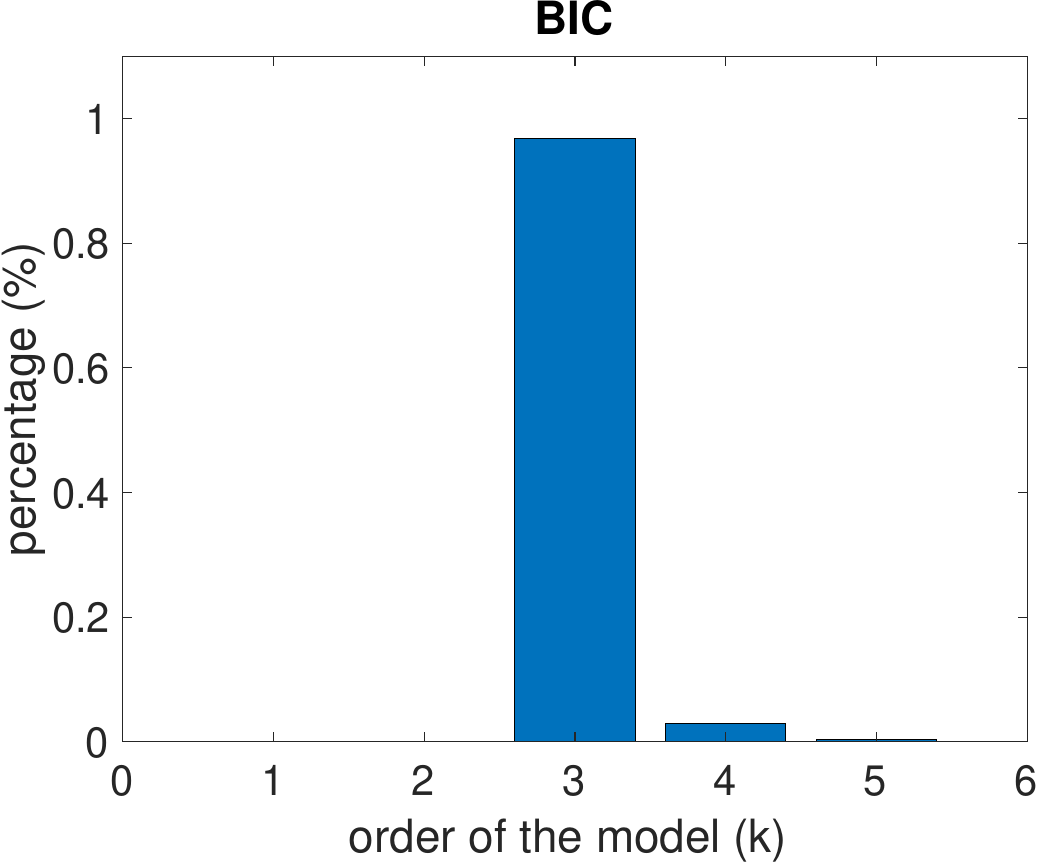}}
\subfigure[]{\includegraphics[width=5.1cm]{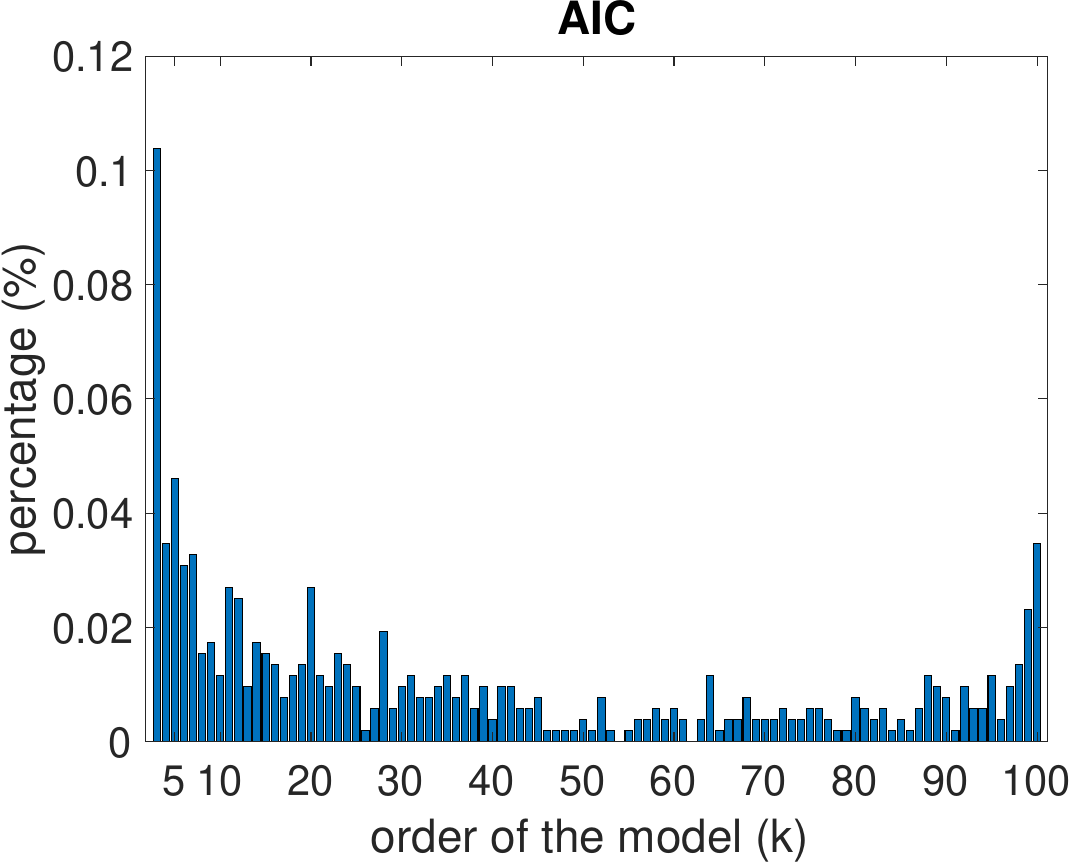}}
\subfigure[]{\includegraphics[width=5.1cm]{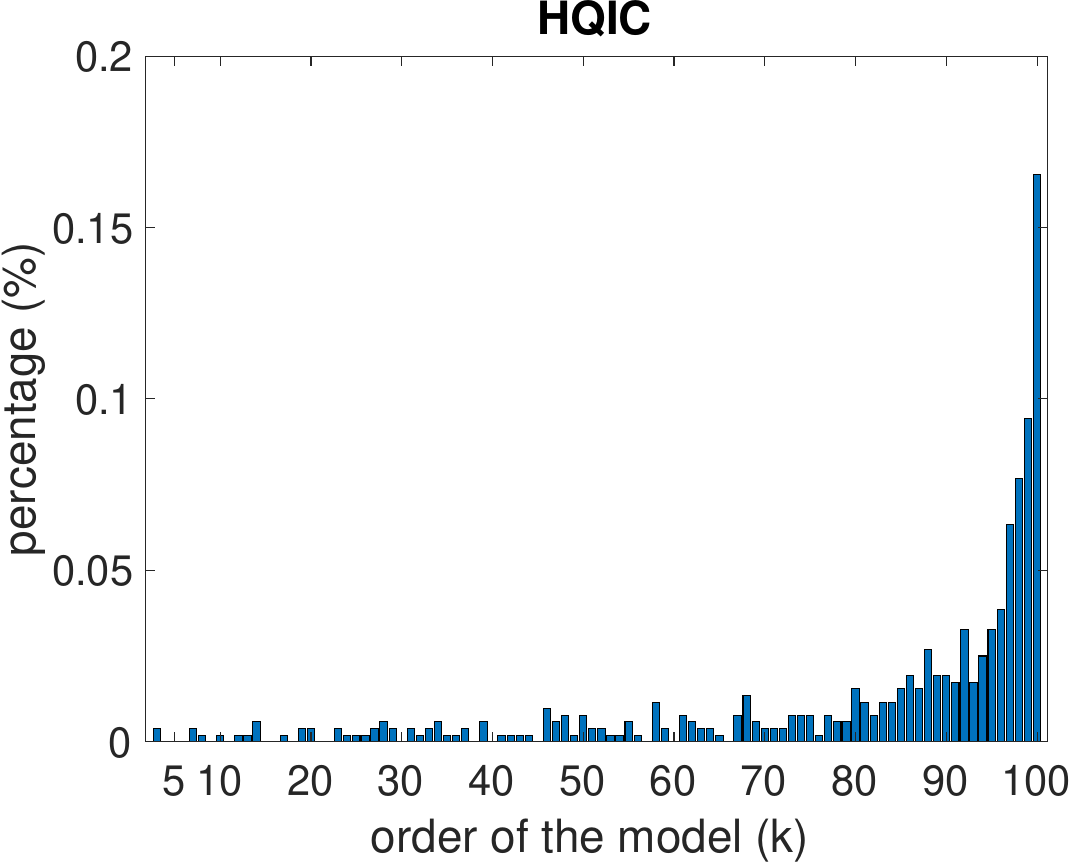}}
}
\caption{Percentages of model order decision by the different methods, in the scenario where the true order is $k=3$,  the standard deviation of the noise is $\sigma_\epsilon=2$, and the number of data $T=200$; {\bf (a)} results of UAED; {\bf (b)} results of  BIC; {\bf (c)} results of  AIC;  {\bf (d)} results of HQIC.}
\label{ExARfig5}
\end{figure}
\begin{figure}[h!]
\centerline{
\subfigure[]{\includegraphics[width=5cm]{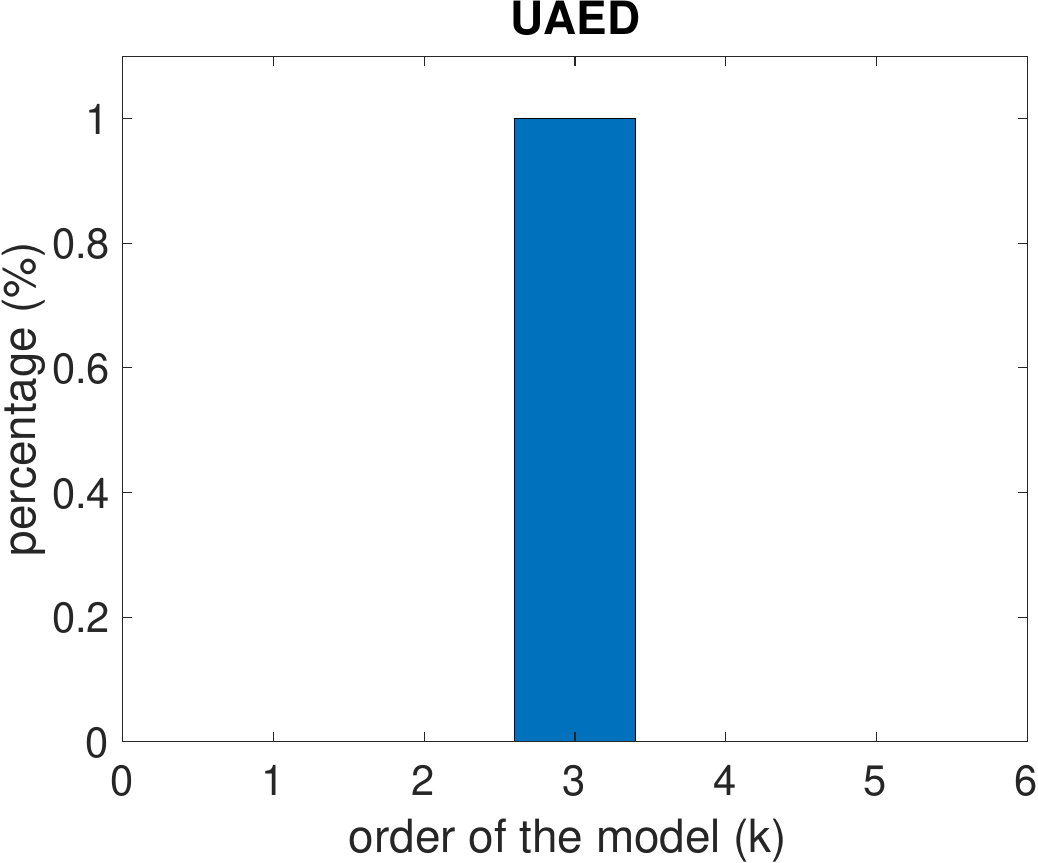}}
\subfigure[]{\includegraphics[width=5cm]{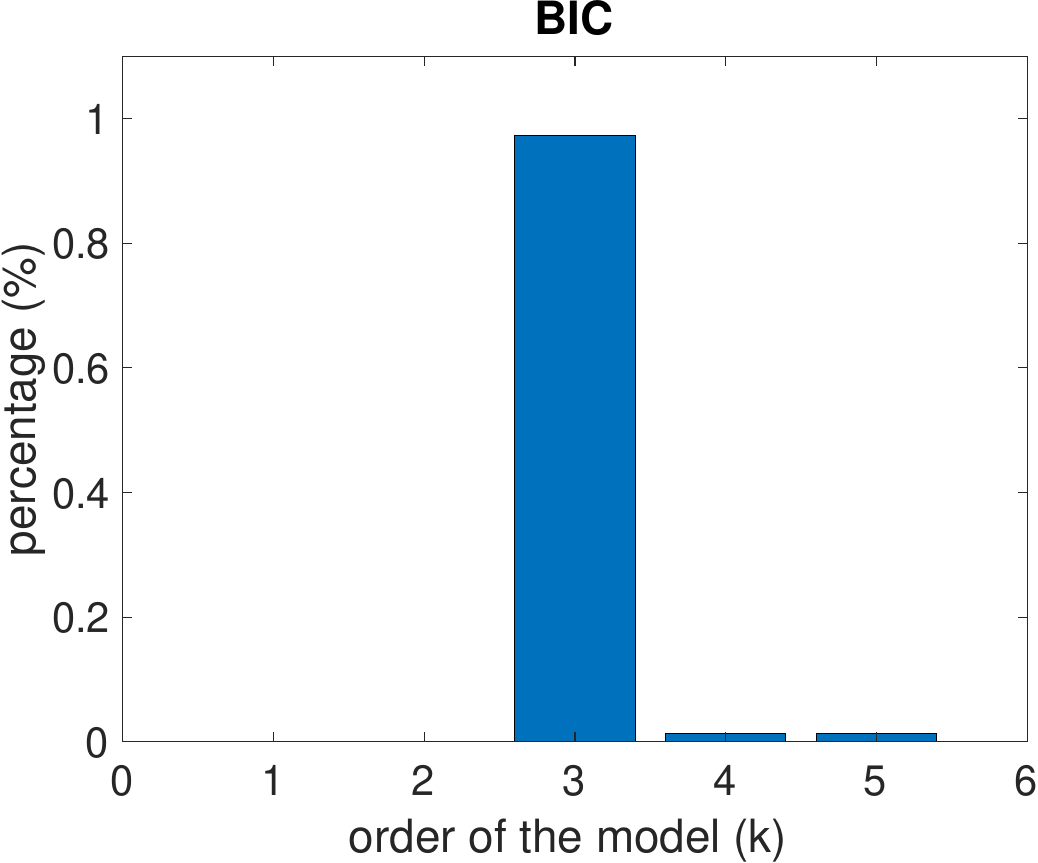}}
\subfigure[]{\includegraphics[width=5.1cm]{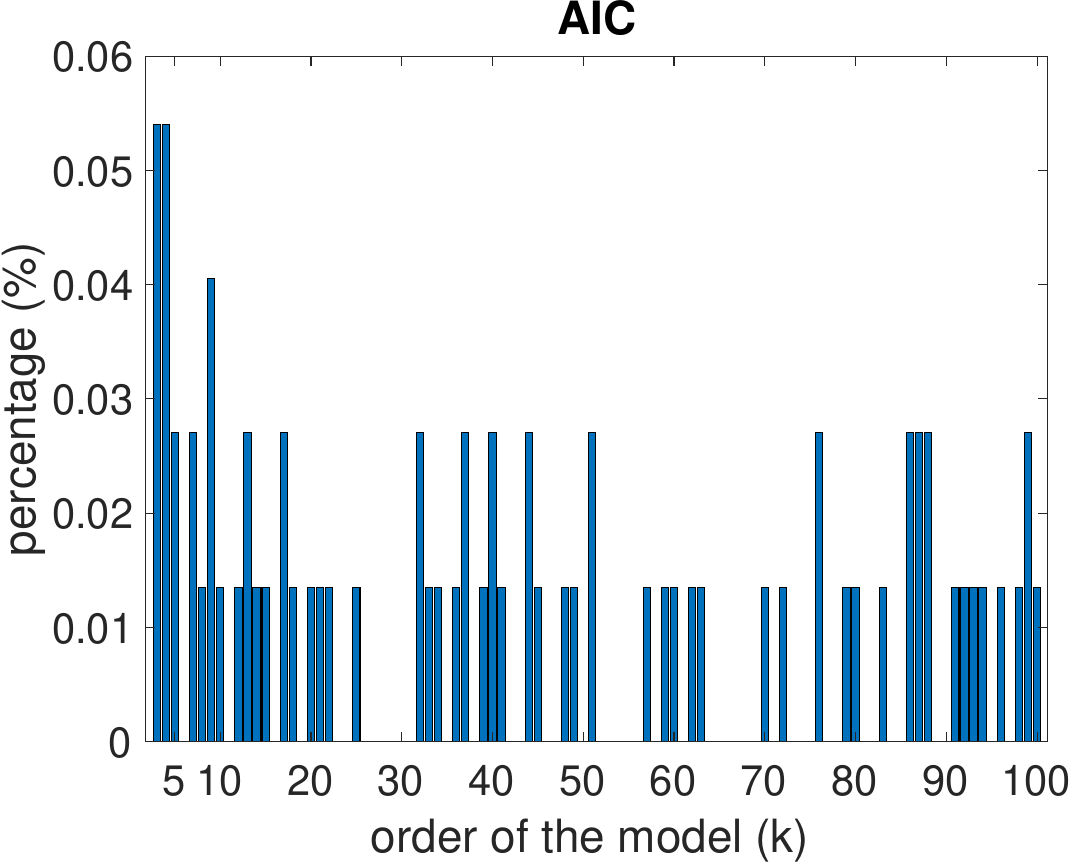}}
\subfigure[]{\includegraphics[width=5.1cm]{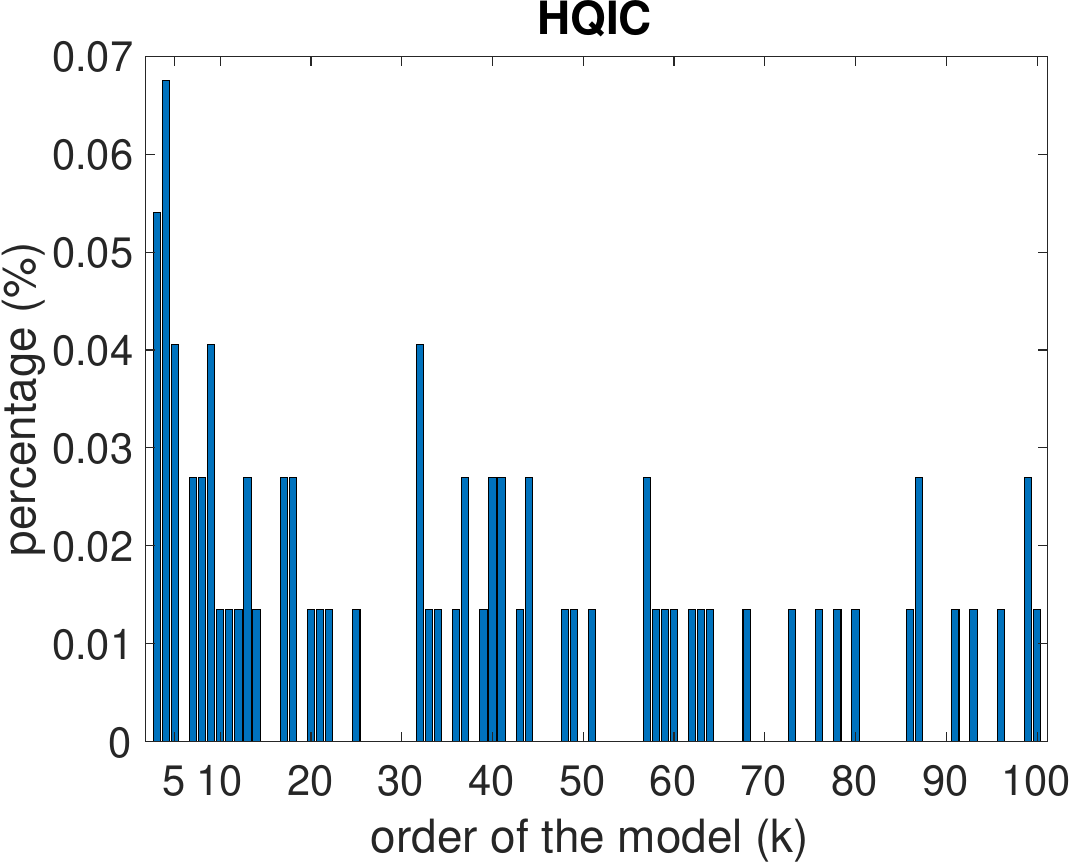}}
}
\caption{Percentages of model order decision by the different methods, in the scenario where the true order is $k=3$,  the standard deviation of the noise is $\sigma_\epsilon=2$, and the number of data $T=2000$; {\bf (a)} results of UAED; {\bf (b)} results of  BIC; {\bf (c)} results of  AIC;  {\bf (d)} results of HQIC.}
\label{ExARfig6}
\end{figure}
\begin{figure}[h!]
\centerline{
\subfigure[]{\includegraphics[width=5cm]{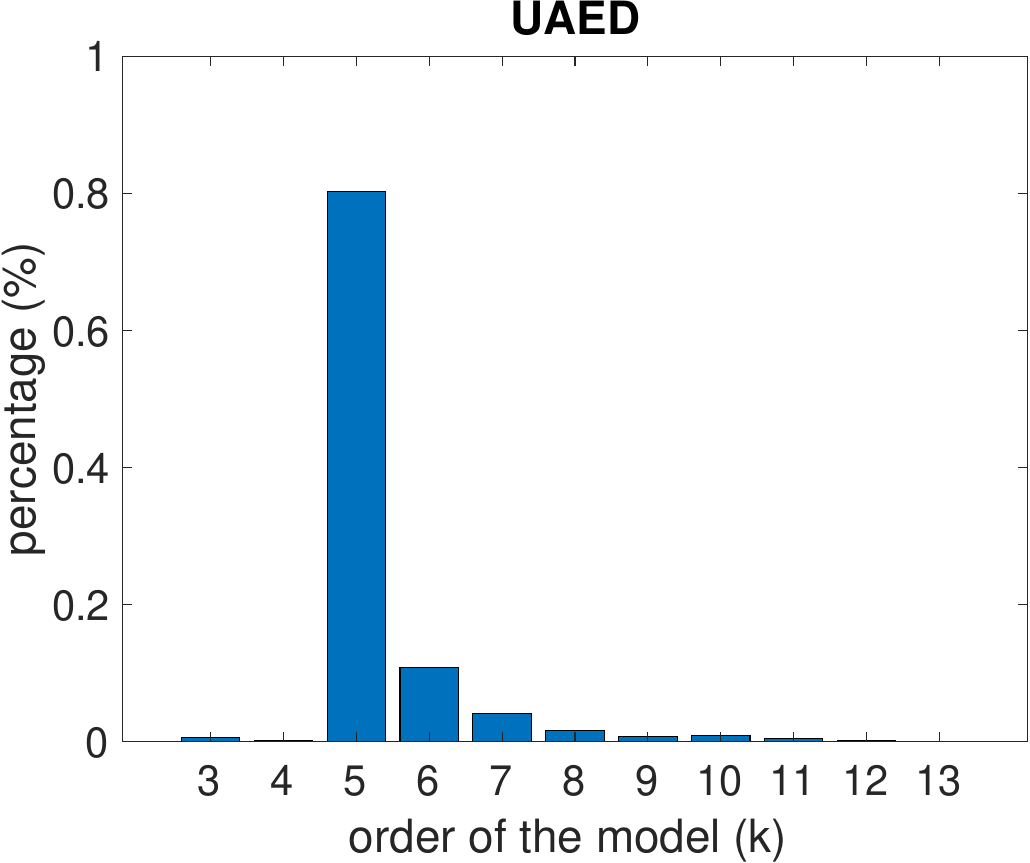}}
\subfigure[]{\includegraphics[width=5cm]{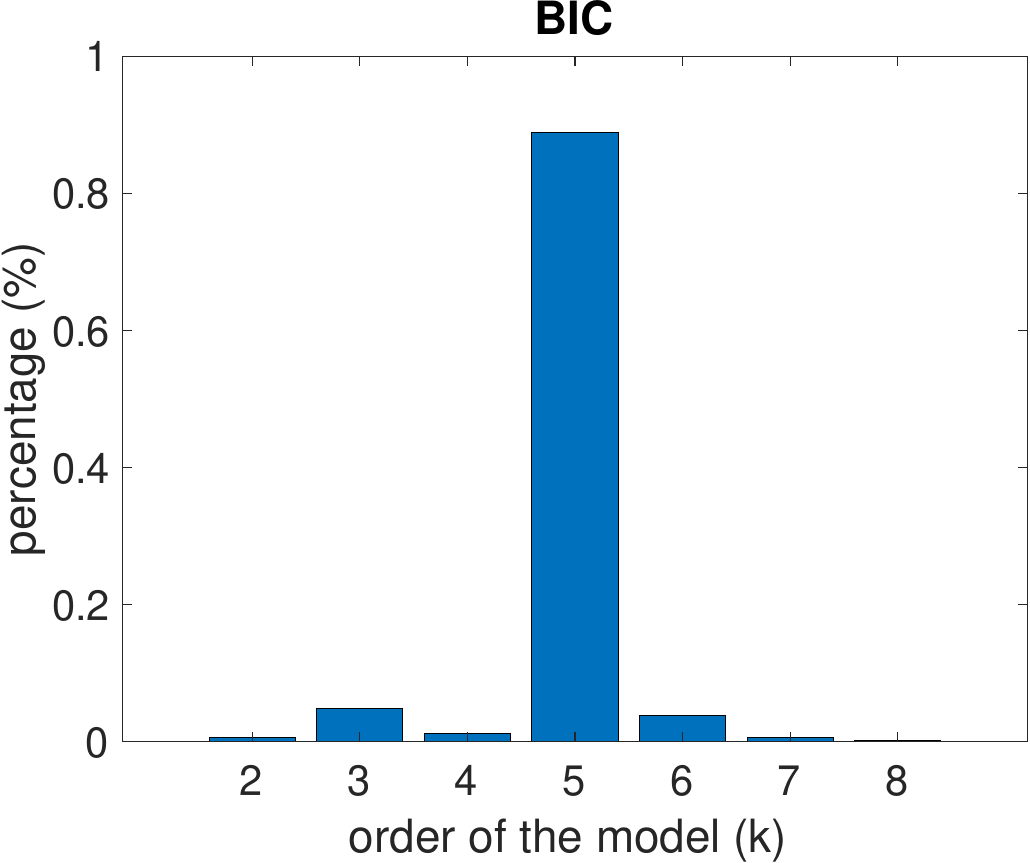}}
\subfigure[]{\includegraphics[width=5.1cm]{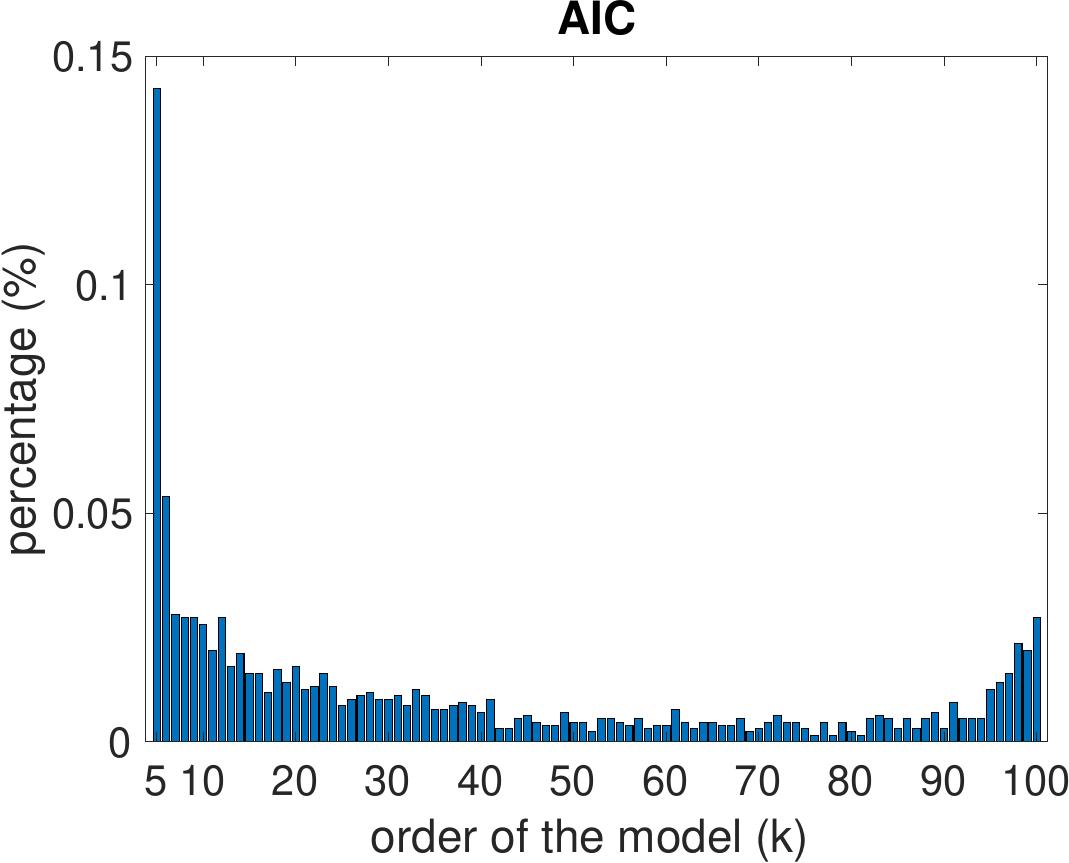}}
\subfigure[]{\includegraphics[width=5.1cm]{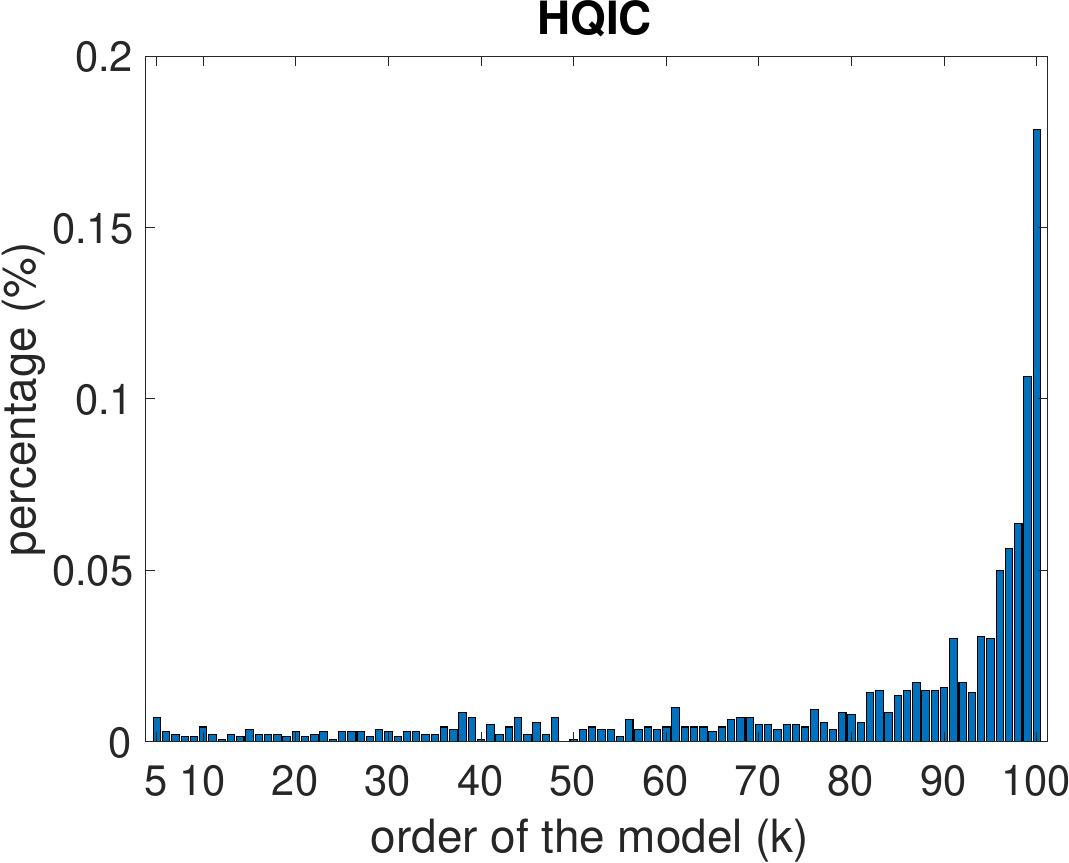}}
}
\caption{Percentages of model order decision by the different methods, in the scenario where the true order is $k=5$,  the standard deviation of the noise is $\sigma_\epsilon=0.5$, and the number of data $T=200$; {\bf (a)} results of UAED; {\bf (b)} results of  BIC; {\bf (c)} results of  AIC;  {\bf (d)} results of HQIC.}
\label{ExARfig7}
\end{figure}
\begin{figure}[h!]
\centerline{
\subfigure[]{\includegraphics[width=5cm]{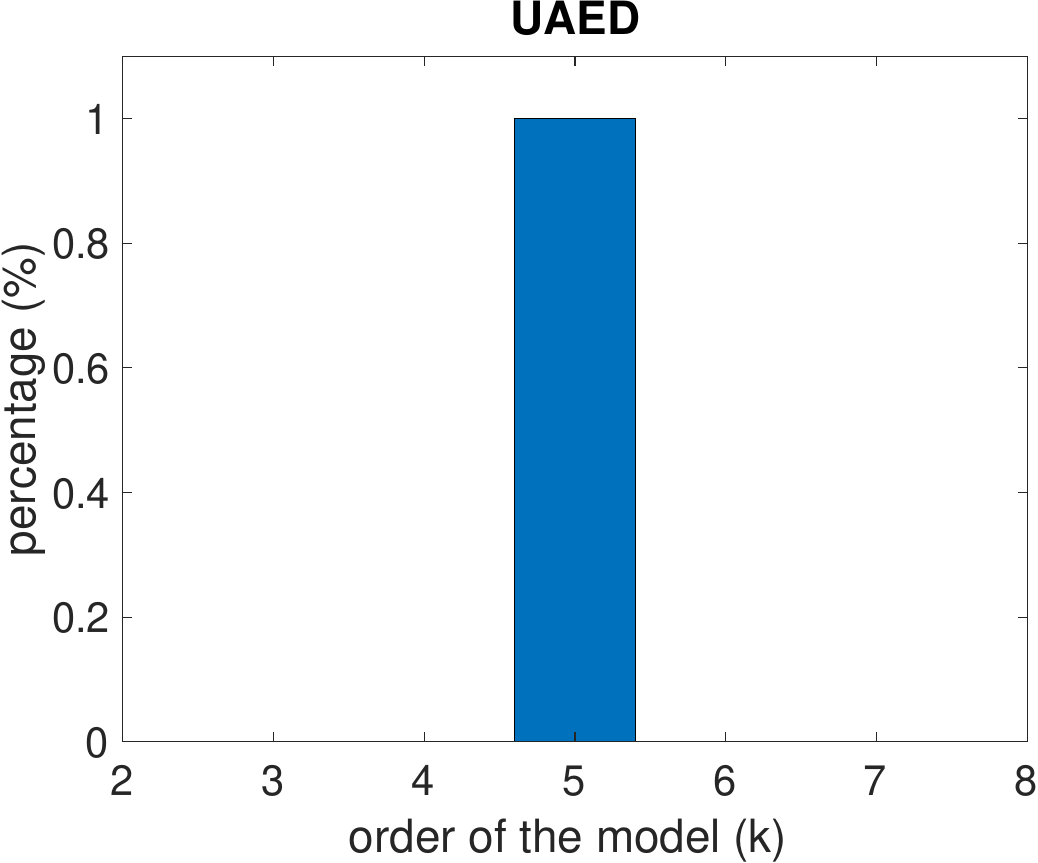}}
\subfigure[]{\includegraphics[width=5cm]{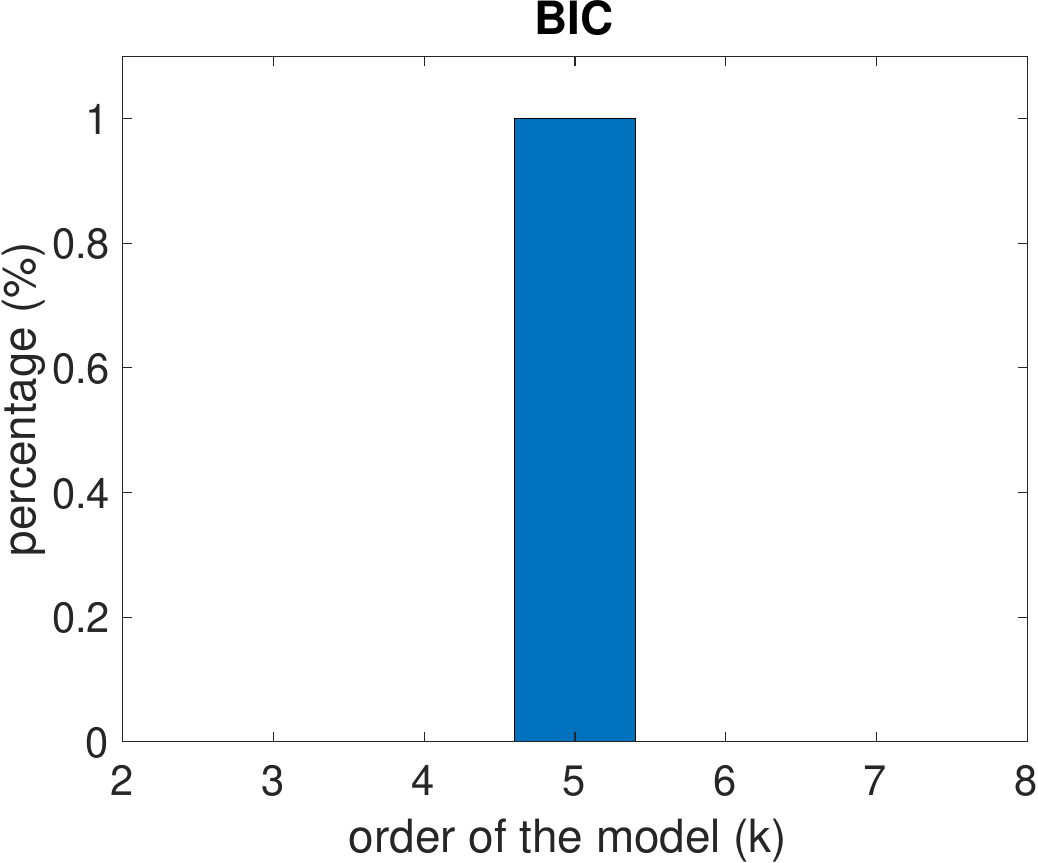}}
\subfigure[]{\includegraphics[width=5.1cm]{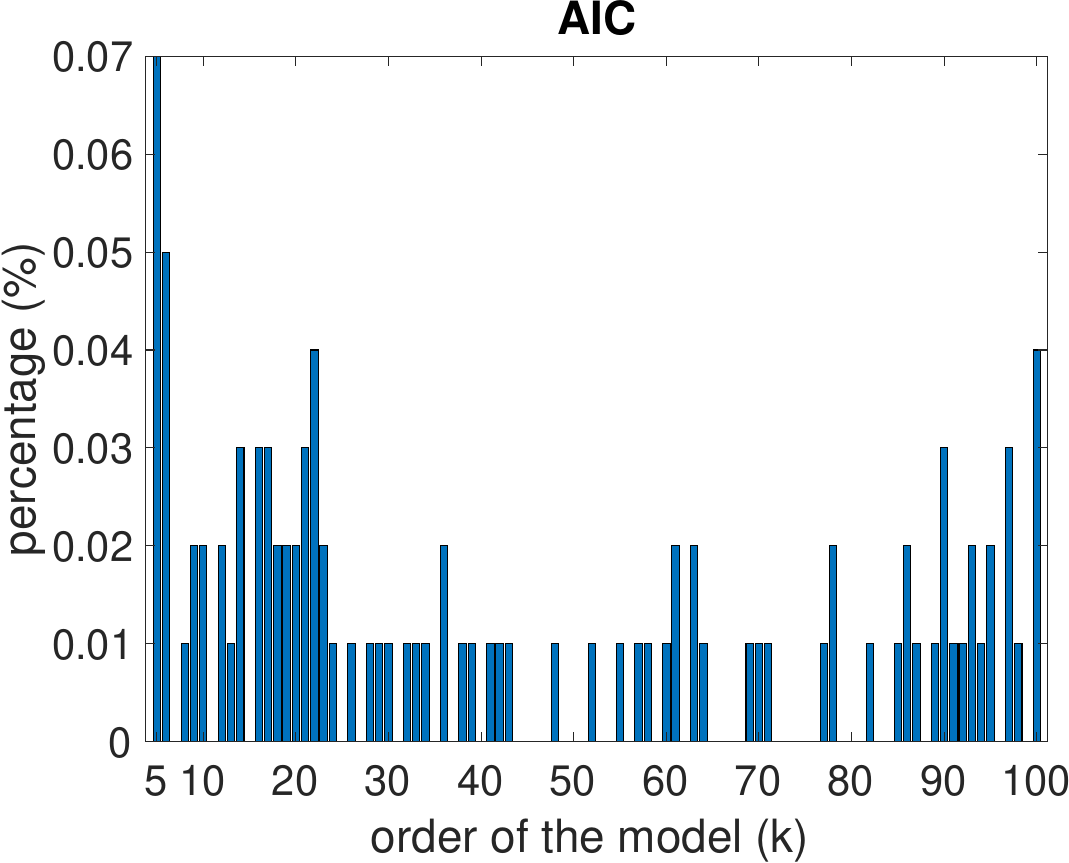}}
\subfigure[]{\includegraphics[width=5.1cm]{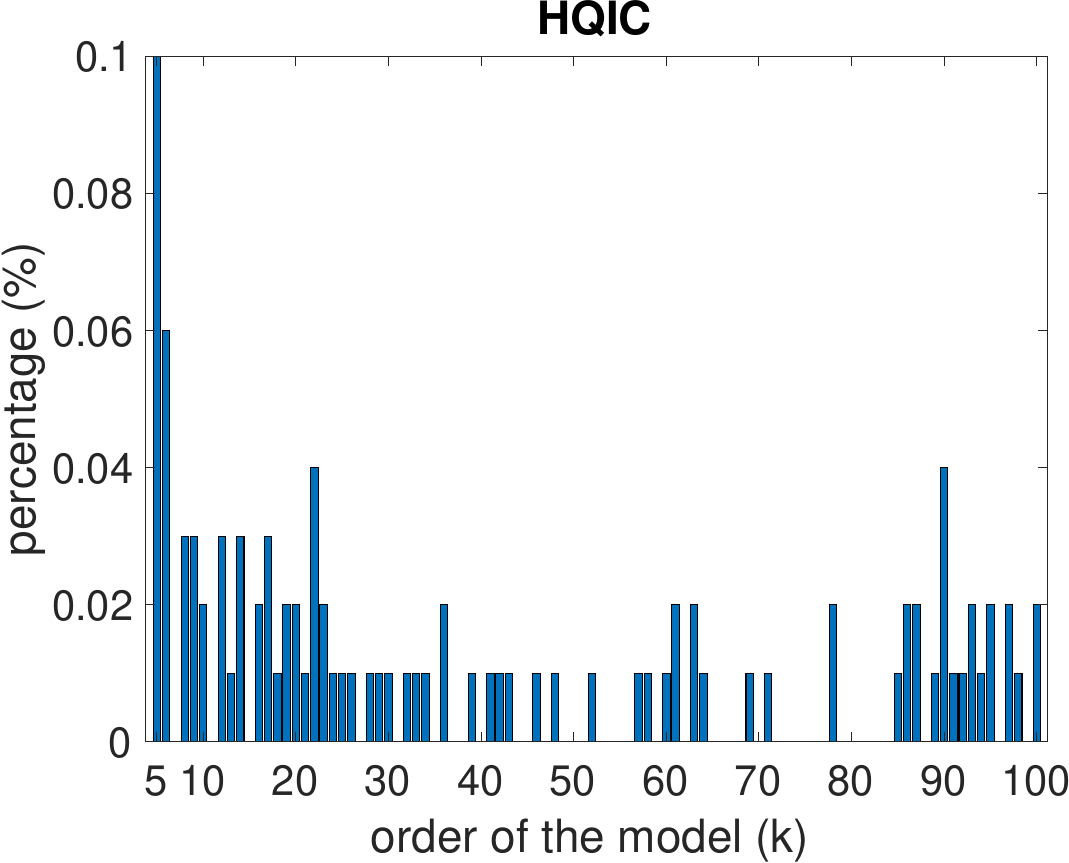}}
}
\caption{Percentages of model order decision by the different methods, in the scenario where the true order is $k=5$,  the standard deviation of the noise is $\sigma_\epsilon=0.5$, and the number of data $T=2000$; {\bf (a)} results of UAED; {\bf (b)} results of  BIC; {\bf (c)} results of  AIC;  {\bf (d)} results of HQIC.}
\label{ExARfig8}
\end{figure}
\begin{figure}[h!]
\centerline{
\subfigure[]{\includegraphics[width=5cm]{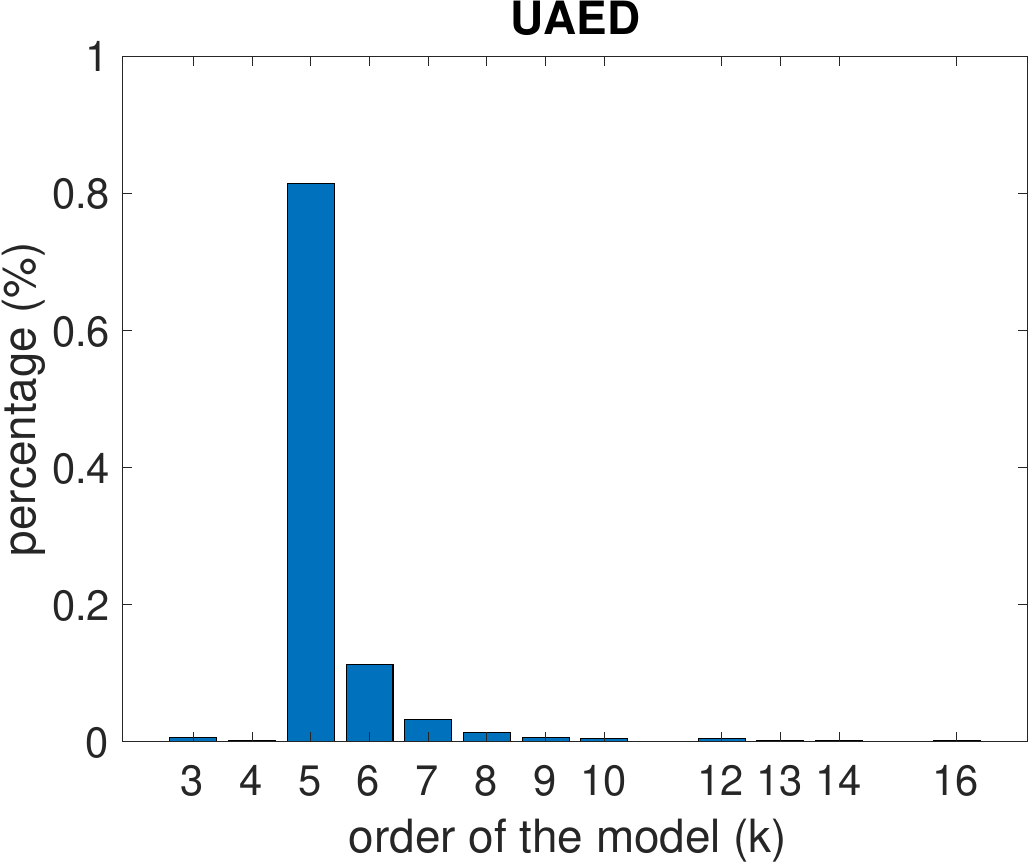}}
\subfigure[]{\includegraphics[width=5cm]{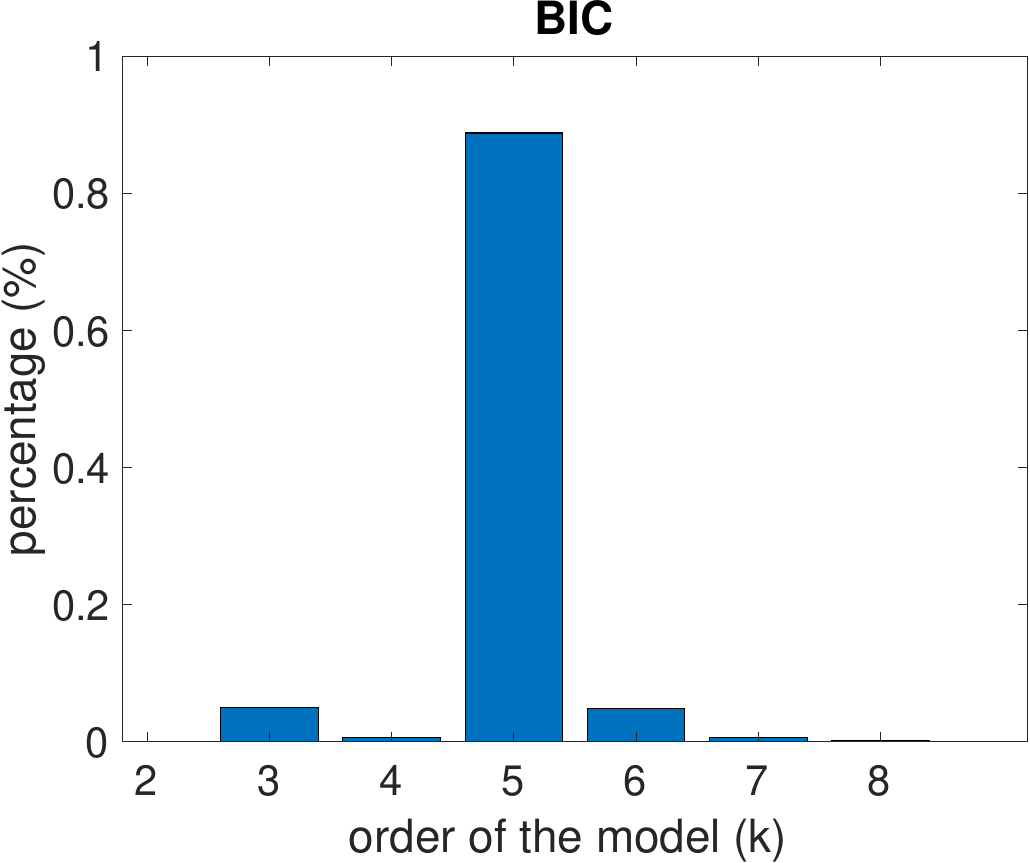}}
\subfigure[]{\includegraphics[width=5.1cm]{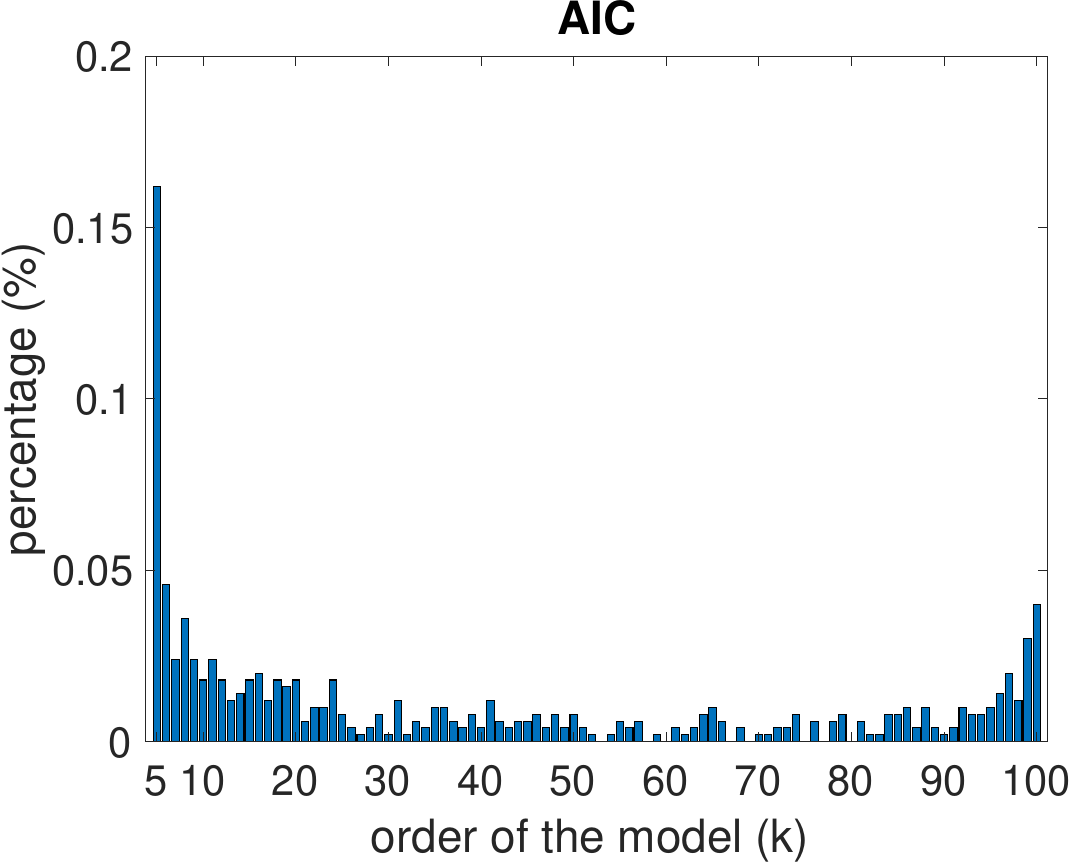}}
\subfigure[]{\includegraphics[width=5.1cm]{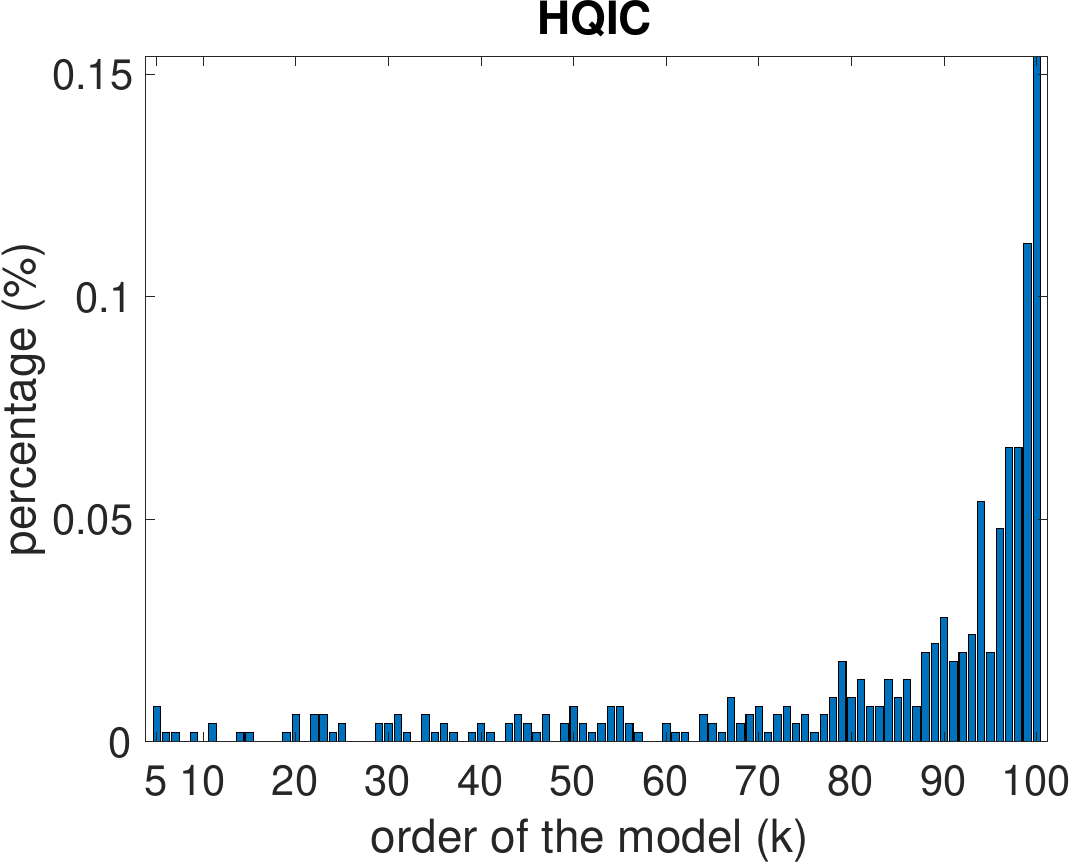}}
}
\caption{Percentages of model order decision by the different methods, in the scenario where the true order is $k=5$,  the standard deviation of the noise is $\sigma_\epsilon=1$, and the number of data $T=200$; {\bf (a)} results of UAED; {\bf (b)} results of  BIC; {\bf (c)} results of  AIC;  {\bf (d)} results of HQIC.}
\label{ExARfig9}
\end{figure}
\begin{figure}[h!]
\centerline{
\subfigure[]{\includegraphics[width=5cm]{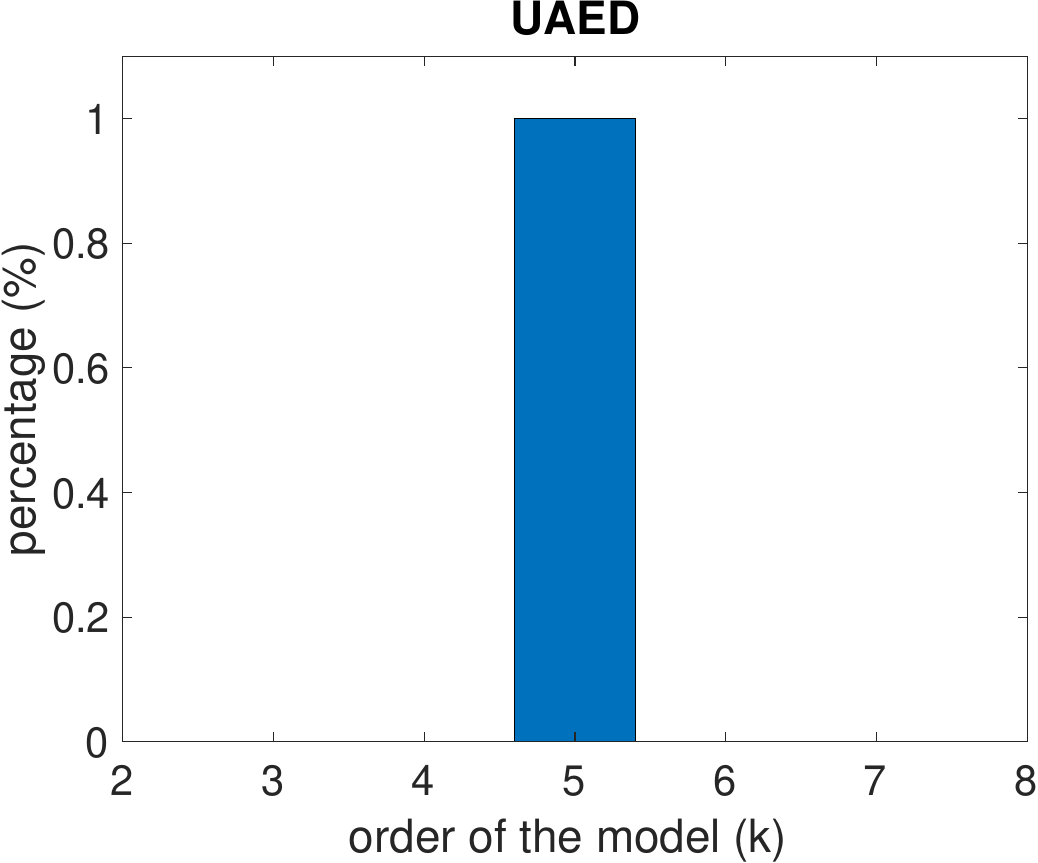}}
\subfigure[]{\includegraphics[width=5cm]{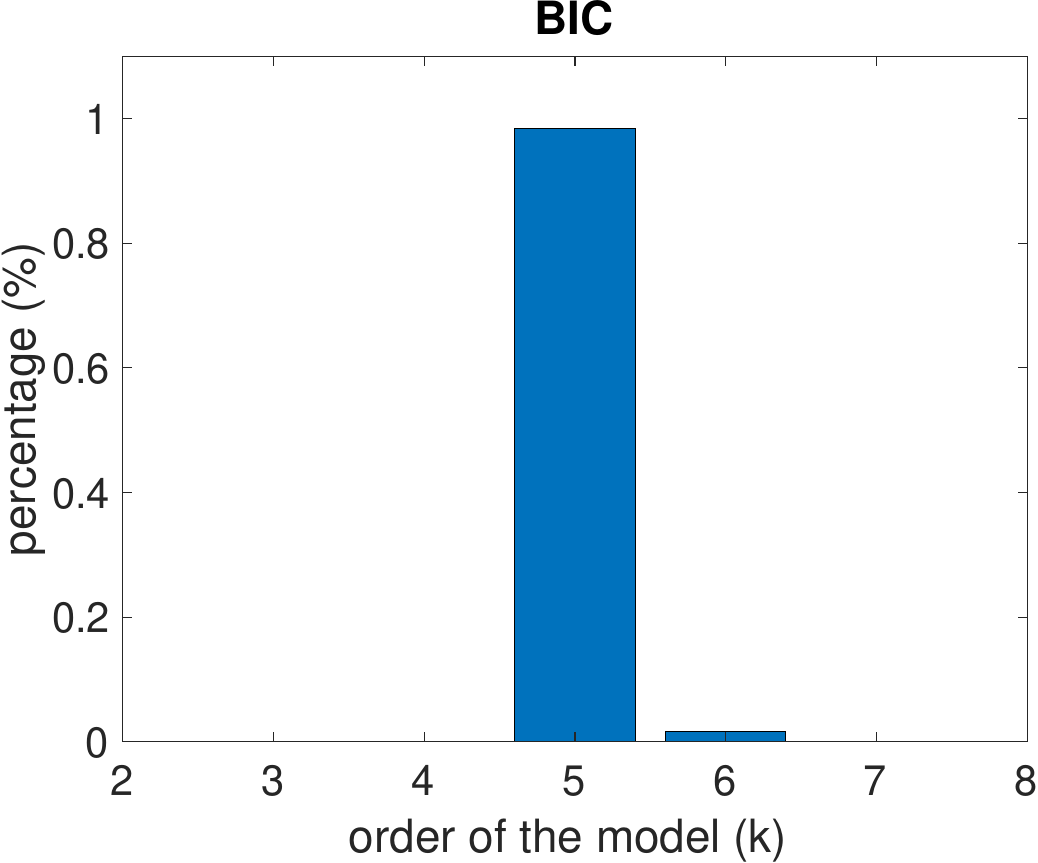}}
\subfigure[]{\includegraphics[width=5.1cm]{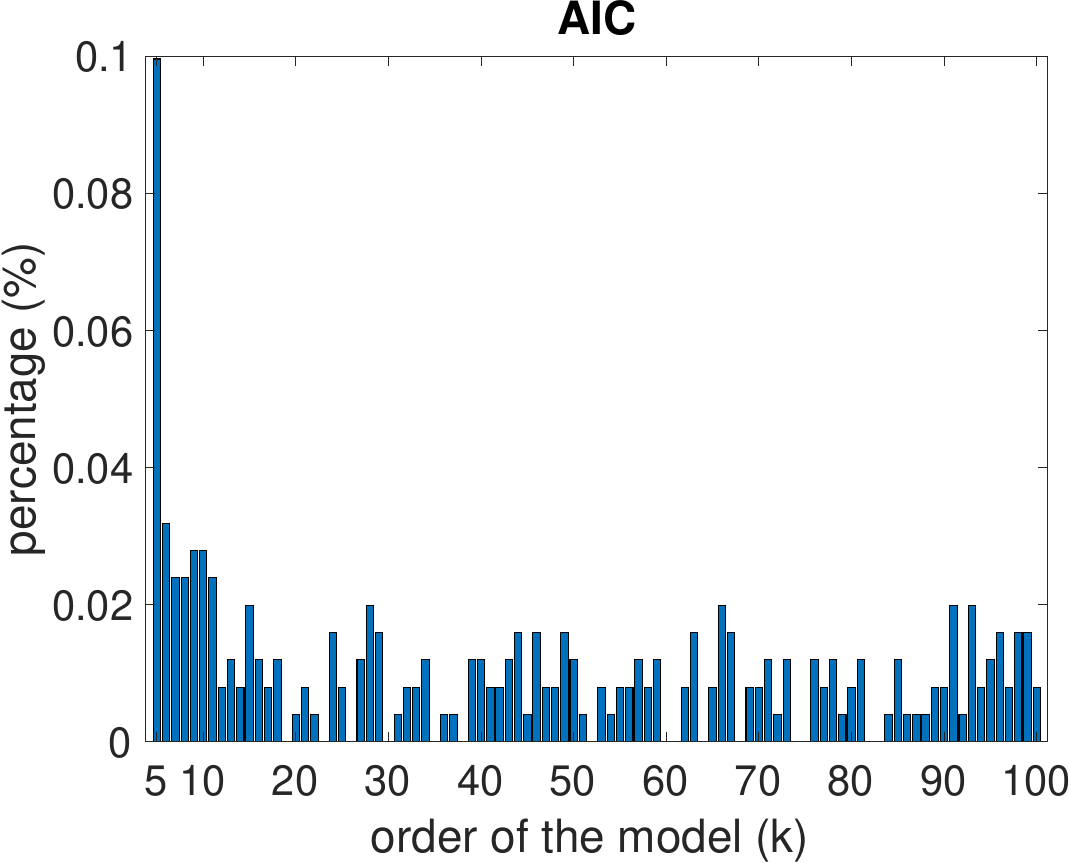}}
\subfigure[]{\includegraphics[width=5.1cm]{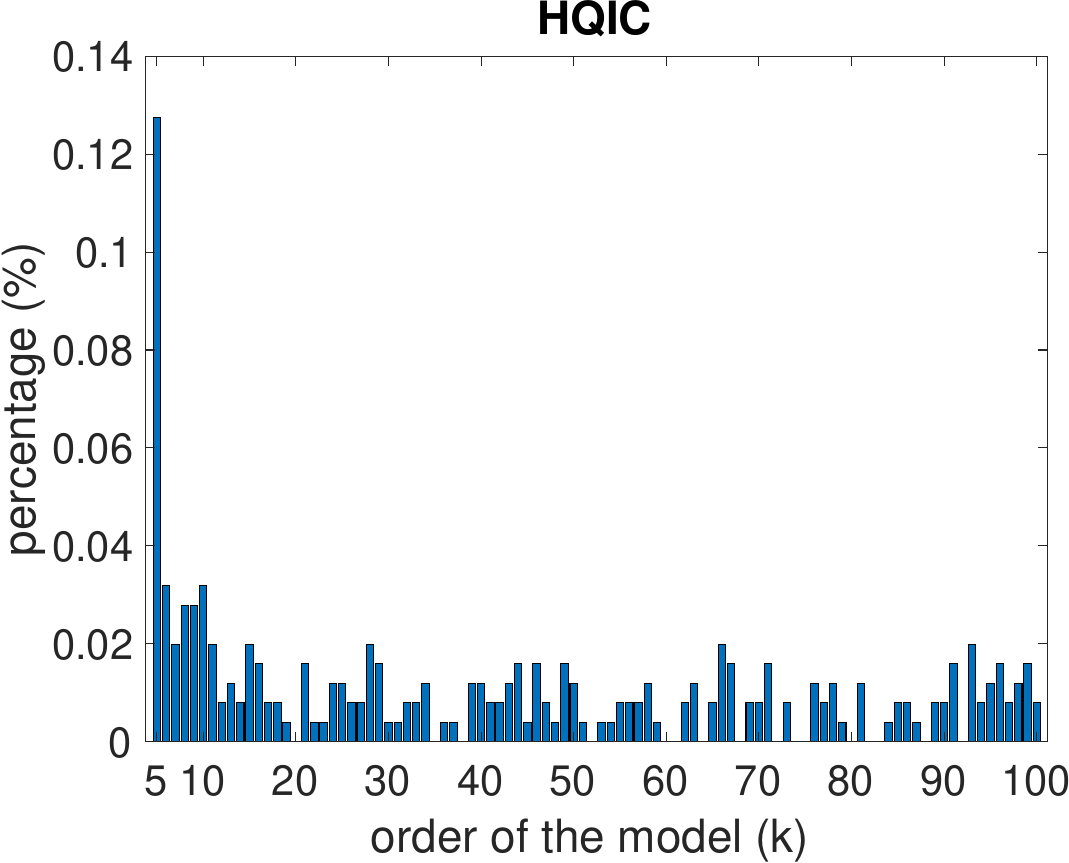}}
}
\caption{Percentages of model order decision by the different methods, in the scenario where the true order is $k=5$,  the standard deviation of the noise is $\sigma_\epsilon=1$, and the number of data $T=2000$; {\bf (a)} results of UAED; {\bf (b)} results of  BIC; {\bf (c)} results of  AIC;  {\bf (d)} results of HQIC.}
\label{ExARfig10}
\end{figure}
\begin{figure}[h!]
\centerline{
\subfigure[]{\includegraphics[width=5cm]{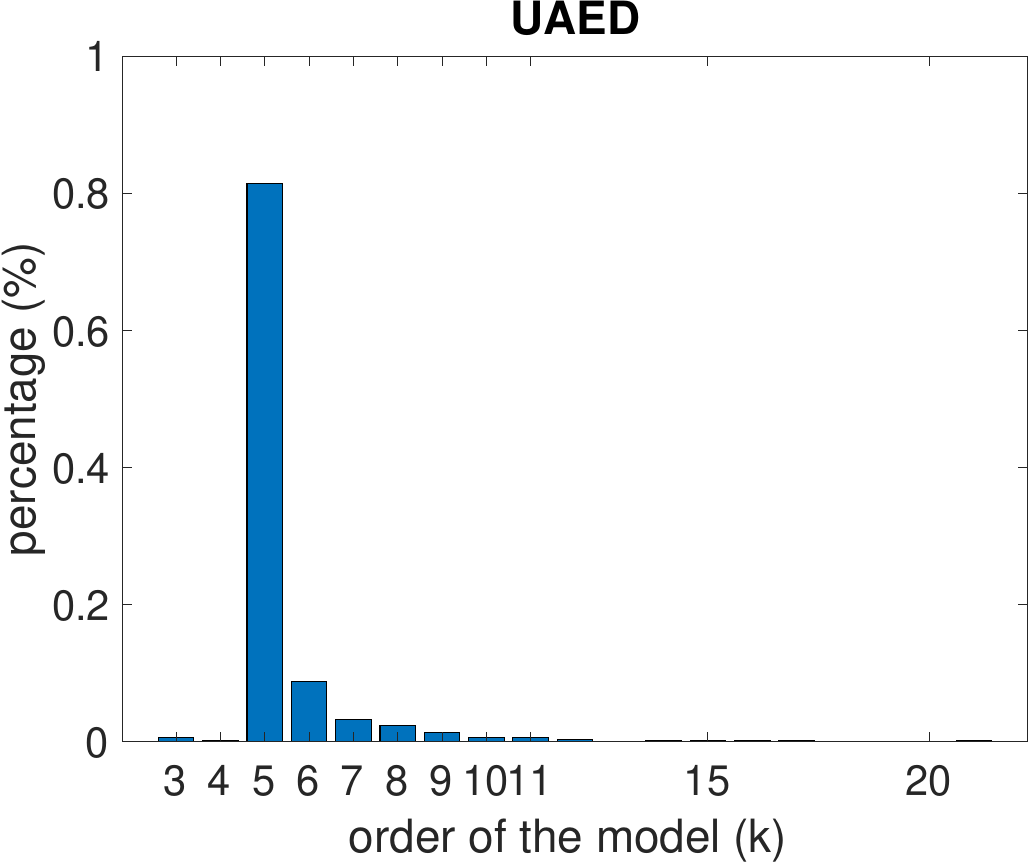}}
\subfigure[]{\includegraphics[width=5cm]{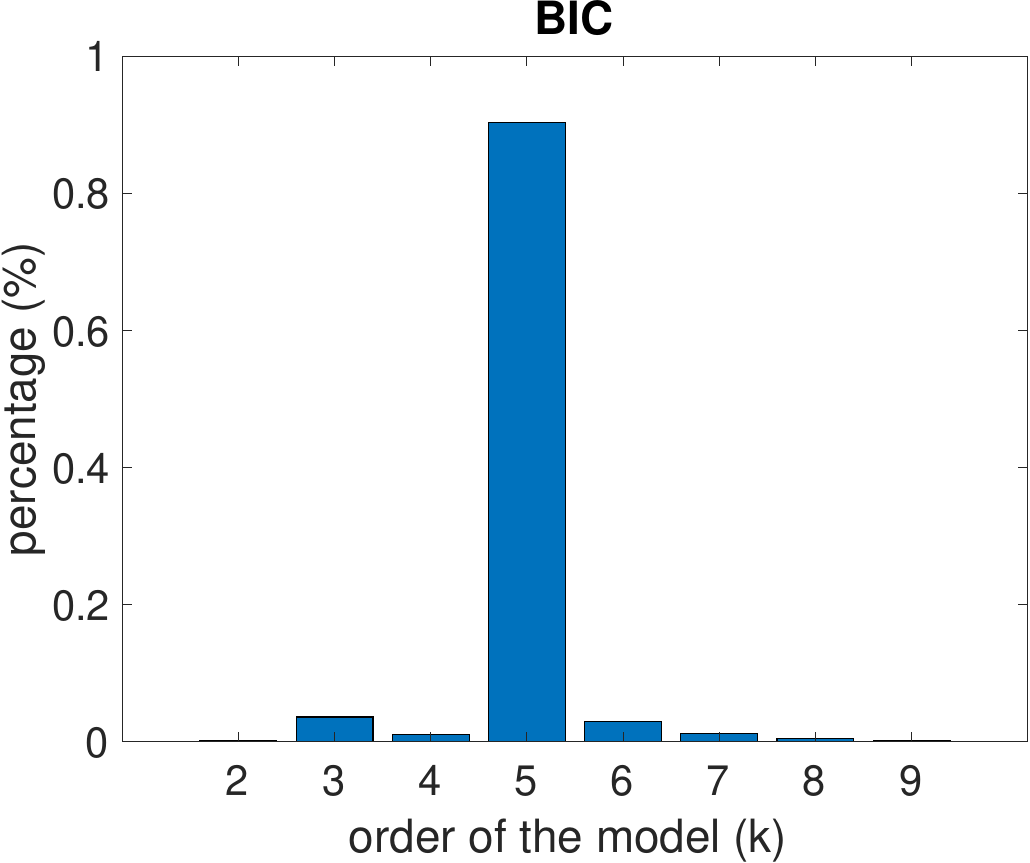}}
\subfigure[]{\includegraphics[width=5.1cm]{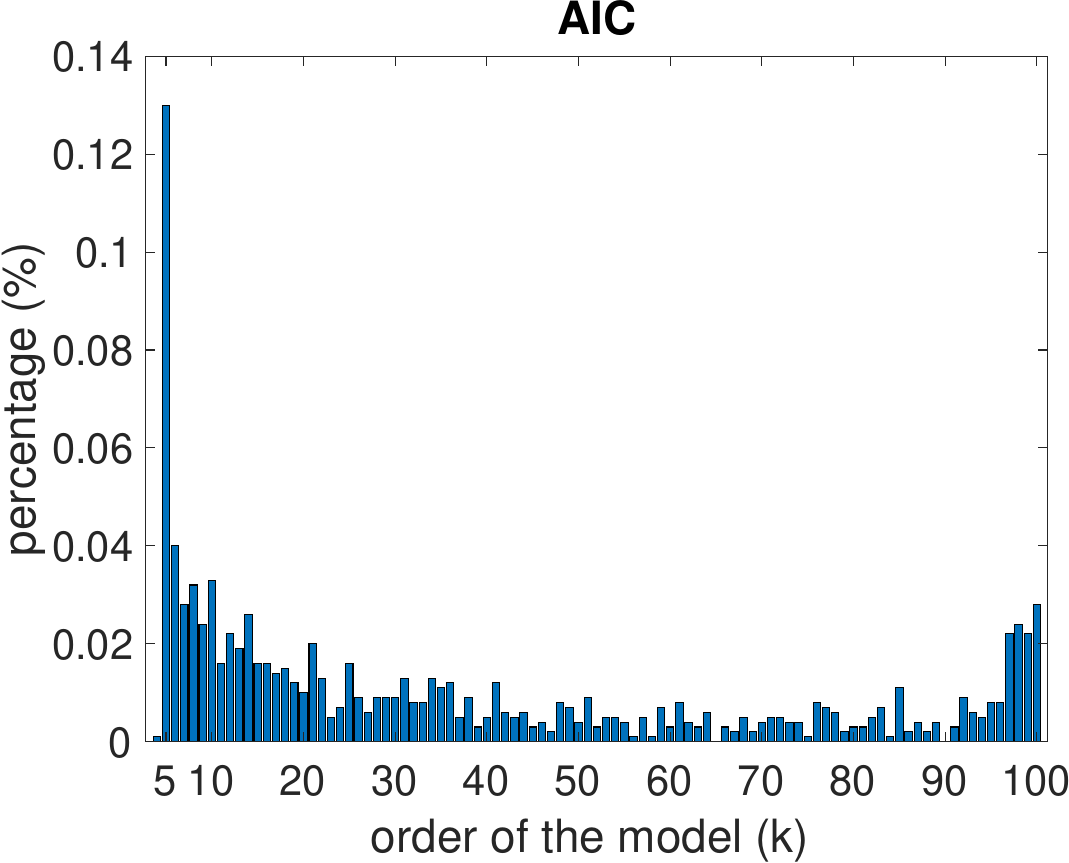}}
\subfigure[]{\includegraphics[width=5.1cm]{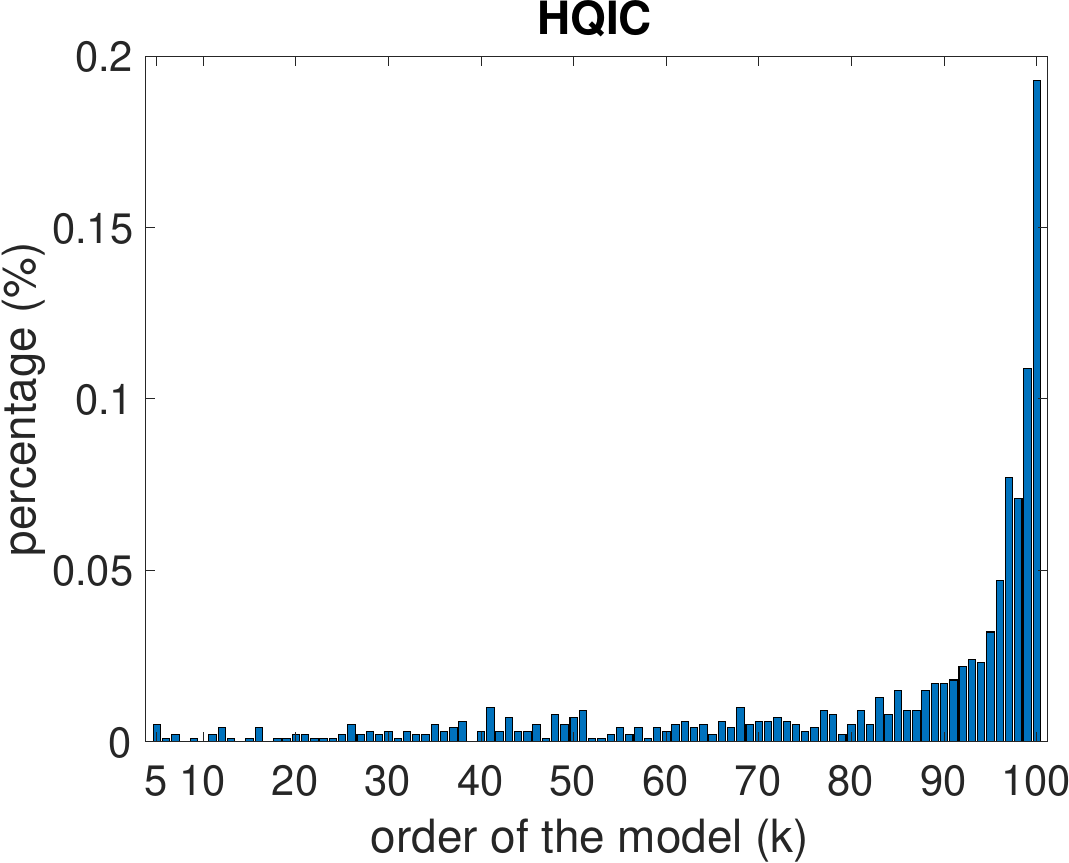}}
}
\caption{Percentages of model order decision by the different methods, in the scenario where the true order is $k=5$,  the standard deviation of the noise is $\sigma_\epsilon=2$, and the number of data $T=200$; {\bf (a)} results of UAED; {\bf (b)} results of  BIC; {\bf (c)} results of  AIC;  {\bf (d)} results of HQIC.}
\label{ExARfig11}
\end{figure}
\begin{figure}[h!]
\centerline{
\subfigure[]{\includegraphics[width=5cm]{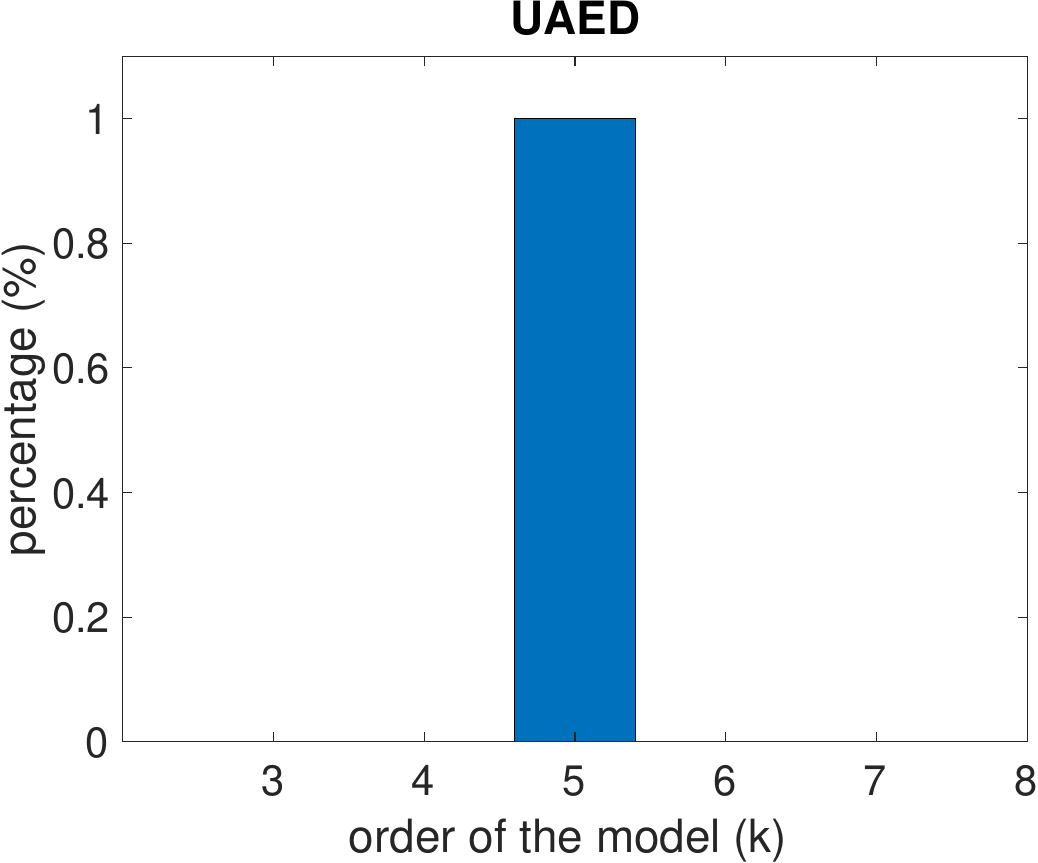}}
\subfigure[]{\includegraphics[width=5cm]{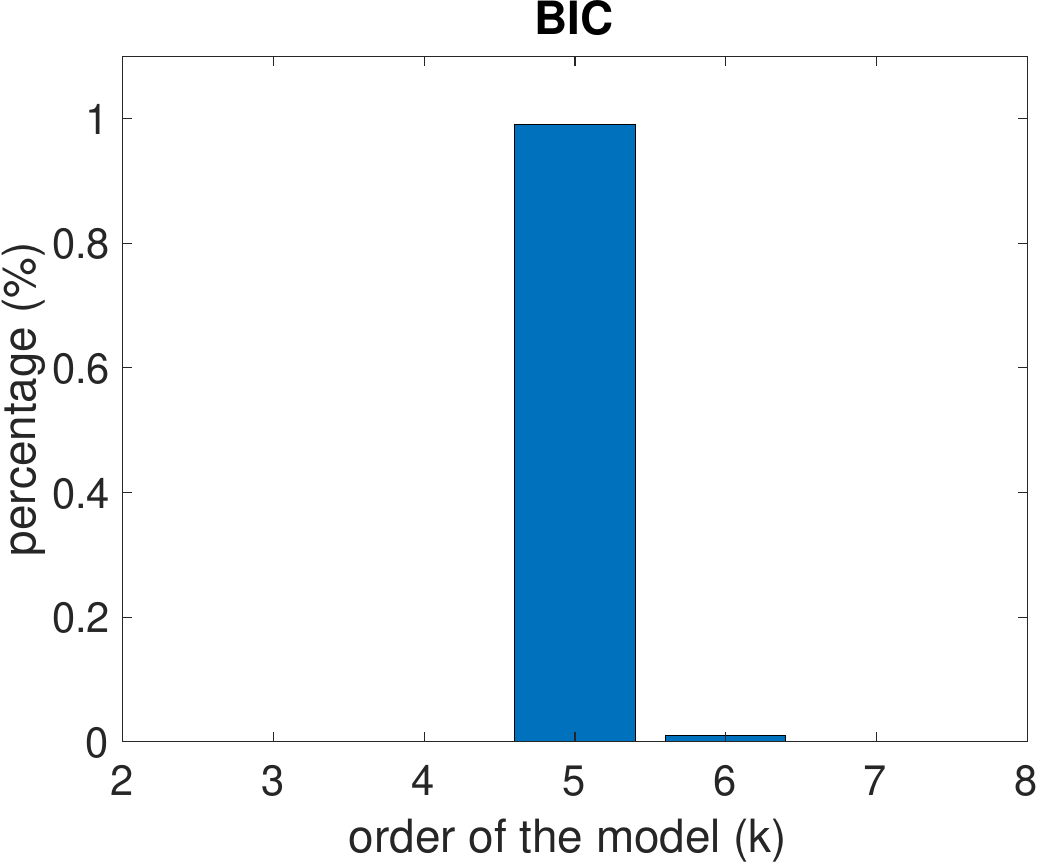}}
\subfigure[]{\includegraphics[width=5.1cm]{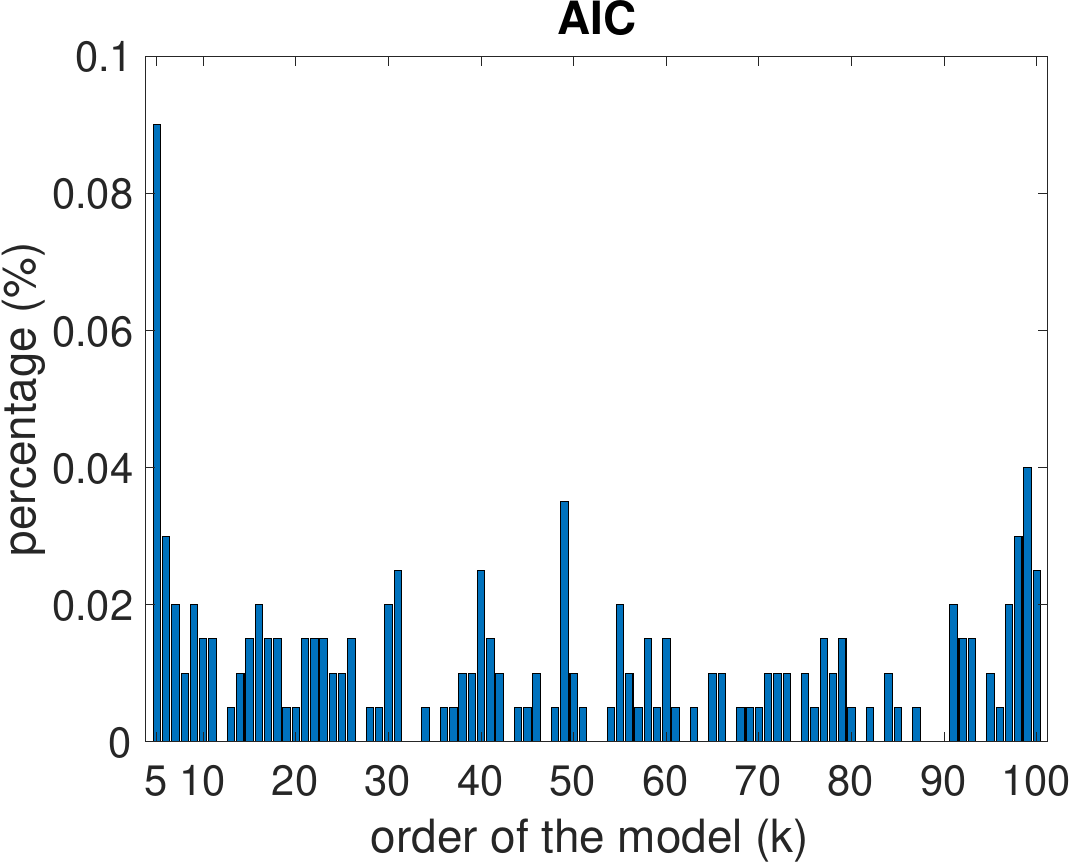}}
\subfigure[]{\includegraphics[width=5.1cm]{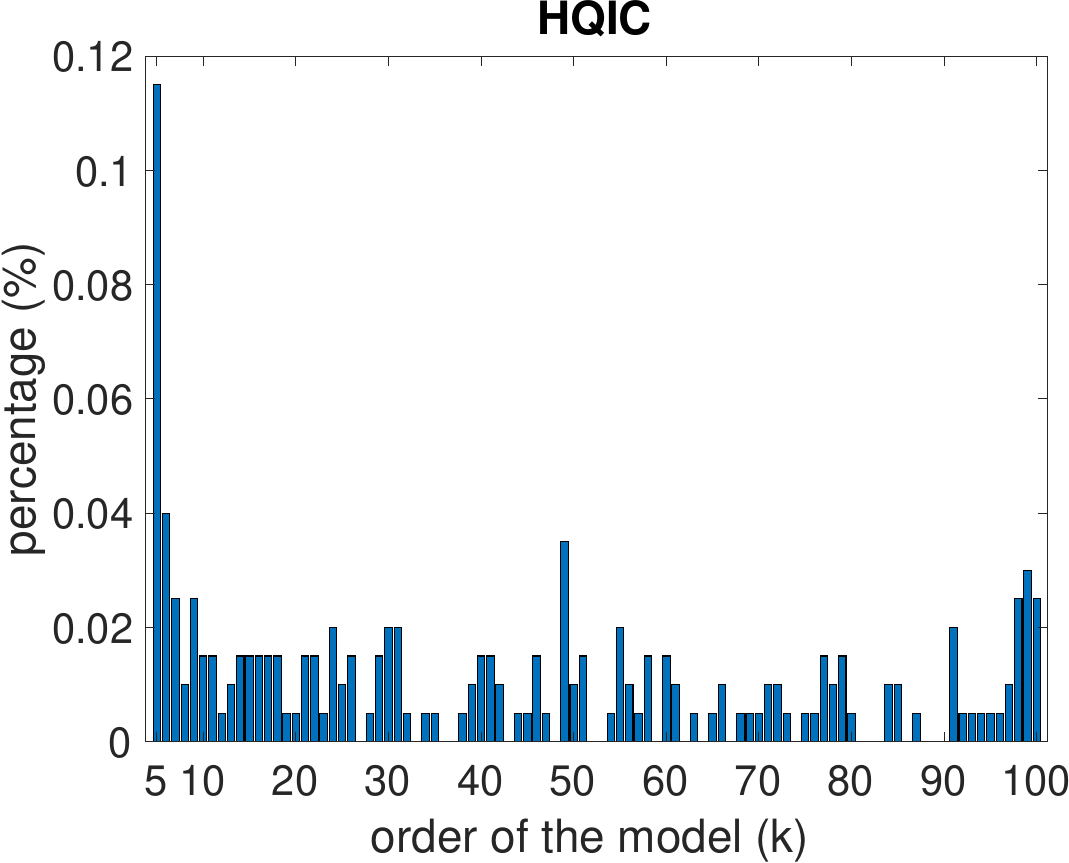}}
}
\caption{Percentages of model order decision by the different methods, in the scenario where the true order is $k=5$,  the standard deviation of the noise is $\sigma_\epsilon=2$, and the number of data $T=2000$; {\bf (a)} results of UAED; {\bf (b)} results of  BIC; {\bf (c)} results of  AIC;  {\bf (d)} results of HQIC.}
\label{ExARfig12}
\end{figure}

\end{document}